\documentclass[aps,prd,floats,floatfix,showpacs,twoside,preprintnumbers,superscriptaddress,nofootinbib]{revtex4}
\usepackage[dvips]{graphicx,pstricks}
\usepackage{amsmath}
\usepackage{amssymb}
\usepackage{epsfig}
\usepackage{color}

\newcommand{\be}{\begin{equation}}
\newcommand{\ee}{\end{equation}}
\newcommand{\ba}{\begin{eqnarray}}
\newcommand{\ea}{\end{eqnarray}}

\def\roughly#1{\mathrel{\raise.3ex\hbox{$#1$\kern-.75em%
\lower1ex\hbox{$\sim$}}}}

\def\slashchar#1{\setbox0=\hbox{$#1$}  
   \dimen0=\wd0     
   \setbox1=\hbox{/} \dimen1=\wd1  
   \ifdim\dimen0>\dimen1   
      \rlap{\hbox to \dimen0{\hfil/\hfil}} 
      #1     
   \else     
      \rlap{\hbox to \dimen1{\hfil$#1$\hfil}} 
      /      
   \fi}

\makeatletter
\def\overbracket#1{\mathop{\vbox{\ialign{##\crcr\noalign{\kern3\p@}
\downbracketfill\crcr\noalign{\kern3\p@\nointerlineskip}
$\hfil\displaystyle{#1}\hfil$\crcr}}}\limits}
\def\underbracket#1{\mathop{\vtop{\ialign{##\crcr
$\hfil\displaystyle{#1}\hfil$\crcr\noalign{\kern3\p@\nointerlineskip}
\upbracketfill\crcr\noalign{\kern3\p@}}}}\limits}
\def\upbracketfill{$\m@th\makesm@sh{\llap{\vrule\@height3\p@\@width.7\p@}}%
\leaders\vrule\@height.7\p@\hfill
\makesm@sh{\rlap{\vrule\@height3\p@\@width.7\p@}}$}
\def\downbracketfill{$\m@th
\makesm@sh{\llap{\vrule\@height.7\p@\@depth2.3\p@\@width.7\p@}}%
\leaders\vrule\@height.7\p@\hfill
\makesm@sh{\rlap{\vrule\@height.7\p@\@depth2.3\p@\@width.7\p@}}$}
\makeatother

\begin{document}

\date{\today}

\title{
Medium induced Lorentz symmetry breaking effects in nonlocal PNJL models}

\author{S.~Beni\' c}
\affiliation{Physics Department, Faculty of Science, University of Zagreb,
Zagreb 10000, Croatia}

\author{D.~Blaschke}
\affiliation{Institut Fizyki Teoretycznej, Uniwersytet Wroc{\l}awski,
50-204 Wroc{\l}aw, Poland}
\affiliation{Bogoliubov Laboratory for Theoretical Physics, JINR Dubna,
141980 Dubna, Russia}
\affiliation{Fakult\"at f\"ur Physik, Universit\"at Bielefeld, 33615 Bielefeld,
Germany}

\author{G.~A.~Contrera}
\affiliation{CONICET, Rivadavia 1917, 1033 Buenos Aires, Argentina}
\affiliation{IFLP, CONICET - Departamento de F\'{i}sica, UNLP, La Plata,
Argentina}
\affiliation{Gravitation, Astrophysics
and Cosmology Group, FCAyG, UNLP, La Plata, Argentina}

\author{D.~Horvati\' c}
\affiliation{Physics Department, Faculty of Science, University of Zagreb,
Zagreb 10000, Croatia}

\begin{abstract}
In this paper we detail the thermodynamics of two flavor nonlocal
Polyakov--Nambu--Jona-Lasinio models for different
parametrizations of the quark interaction regulators.
The structure of the model is upgraded in order to allow for
terms in the quark selfenergy which violate Lorentz invariance
due to the presence of the medium.
We examine the critical properties, the phase diagram as well as
the equation of state.
Furthermore, some aspects of the Mott effect
for pions and sigma mesons are discussed
explicitly within a nonlocal Polyakov--Nambu--Jona-Lasinio model.
In particular, we continued the meson polarization function
in the complex energy plane
and under certain approximations, we were able to extract the 
imaginary part as a function of the meson energy.
We were not able to calculate the dynamical meson mass, and therefore 
resorted to a technical study of the temperature dependence of 
the meson width by replacing the meson energy with the 
temperature dependent spatial meson mass.
Our results show that while the temperature behavior 
of the meson widths is qualitatively the same
for a wide class of covariant regulators, the special
case where the nonlocal interactions are introduced via the
instanton liquid model singles out with a drastically different behavior.
\end{abstract}
\pacs{11.10.St, 05.70.Jk, 12.39.Ki, 11.30.Rd, 11.10.Wx}
\maketitle

\section{Introduction}

Temperatures and densities in heavy ion collisions are well above
the point where hadrons maintain their identity.
Experimental data from RHIC and LHC provide strong evidence
that beyond a certain temperature,
low-energy quantum chromodynamics (QCD)
forms a strongly coupled quark-gluon plasma (QGP) phase
\cite{Shuryak:2004cy,Wiedemann:2012py},
behaving almost like a perfect fluid
of deconfined quark and gluon degrees of freedom.
Future facilities like Nuclotron-based Ion Collider
fAcility (NICA) at JINR and Facility for Antiproton and Ion
Research (FAIR) at GSI will complement these results by studying the region of
extreme densities, thereby allowing a detailed account
on the whole QCD phase diagram \cite{Sorin:2011zz,Bleicher:2011sd}.
Recent reviews on the phase diagram are given, {\it e.g.}, in
Refs.~\cite{Fukushima:2010bq,Fukushima:2011jc}.

A strongly interacting theory can be fully addressed
in lattice simulations.
At present, thermodynamic properties of lattice QCD
can be calculated for physical quark masses
(for latest results of the Wuppertal-Budapest group,
see \cite{Borsanyi:2012rr}).
This leads to the important result, already observed several years ago
\cite{Karsch:2003zq,Karsch:2003vd}, that the lattice data below
and including the pseudocritical temperature is
described
by the hadron resonance gas model \cite{Hagedorn:1965st}.
These findings are now well established \cite{Borsanyi:2010bp}.

Due to the sign problem, lattice calculations are still restricted
to a narrow range of finite baryon number chemical potential
dictated by the convergence radius of Taylor expansion techniques at $\mu=0$.
On the other hand, in continuum studies, concentrated on the low-energy chiral
quark sector, a tremendous amount of work has been
accomplished in exploring the whole QCD phase diagram.
These studies
can be roughly separated into classes ranging from the Nambu-Jona Lasinio (NJL)
model with local quark interactions \cite{Nambu:1961tp,Nambu:1961fr}
(see the reviews \cite{Klimt:1989pm,Klevansky:1992qe,Hatsuda:1994pi,
Buballa:2003qv}
for application to quark matter), to the more fundamental approach to
QCD making use of the tower of integral equations for the
$n$-point functions of Euclidean QCD,
the Dyson-Schwinger equation (DSE) approach
\cite{Roberts:2000aa,Alkofer:2000wg,Fischer:2006ub}.
Quark DSEs usually operate on the level of modelling
an effective gluon propagator for describing the nonperturbative
interaction between quarks and neglecting the ghosts (global color model, see
\cite{Tandy:1997qf}), although a more
complete approach is also being developed, see {\it e.g.},
Ref.~\cite{Fischer:2009wc}.

A \textit{separable} form of the quark-quark interaction
\cite{Schmidt:1994di,Bowler:1994ir,Plant:1997jr,Blaschke:1999ab,Blaschke:2000gd}
bridges the gap between the two approaches: NJL and DSE, giving rise to a
nonlocal NJL (nl-NJL) model
\cite{General:2000zx,GomezDumm:2001fz,GomezDumm:2004sr,Blaschke:2007ce}.
With this development, the quark propagator entails a dynamical mass and wave
function renormalization as is well known from lattice QCD studies, see
\cite{Parappilly:2005ei}.
As additional effect, poles of the quark propagator can be absent from the real
axes \cite{Bowler:1994ir,Plant:1997jr,Blaschke:1998gk}.
It is well known that appearance of, \textit{e. g.},
complex conjugate mass poles (CCMPs) in the propagator
provides a sufficient criteria for confinement
\cite{Bowler:1994ir,Plant:1997jr,Blaschke:1998gk,Roberts:2000aa,
Alkofer:2000wg,Bhagwat:2002tx,Dudal:2008sp,Dudal:2013vha}.
Furthermore, nonlocal models do not require additional cutoffs
\cite{Blaschke:1995gr} and
find no problem in treating anomalies \cite{RuizArriola:1998zi}.
An alternative way to introduce the non-locality is 
inspired by the Instanton Liquid
Model (ILM) \cite{Bowler:1994ir,Plant:1997jr,Buballa:1992sz,
GomezDumm:2006vz,Ripka:1997zb,Schafer:1996wv}.

Recently, the nl-NJL model
was generalized by coupling its chiral quark sector
to the Polyakov loop (PL) variable with an appropriate model for the PL
potential
\cite{Blaschke:2007np,Contrera:2007wu,Hell:2008cc,
Noguera:2008cm,Hell:2009by,
Contrera:2010kz,Horvatic:2010md,Radzhabov:2010dd,Hell:2011ic}.
The most advanced of these nonlocal 
Polyakov--Nambu--Jona-Lasinio (nl-PNJL) approaches address both, scalar and vector
quark selfenergies, like the QCD DSEs do.
It has been demonstrated that these approaches can be embedded in a
scheme which aims towards a first principle derivation of a low-energy QCD
description capable of addressing both, confinement and chiral
symmetry-breaking crossover transitions \cite{Kondo:2010ts}.

The field-theoretic
formulation of such nl-PNJL models
provides a natural starting point for developing them further beyond the
mean field level to address in particular mesonic correlations\footnote{
The description of diquark \cite{GomezDumm:2005hy} and baryonic
\cite{Rezaeian:2005nm} correlations in matter have so far been developed to
the level of the nl-NJL approach without coupling them to the PL.}
\cite{Blaschke:2007np,Hell:2008cc,Hell:2009by,Radzhabov:2010dd}.
The effective mesonic action obtained by integrating out the quark
degrees of freedom reveals its coupling constants as nonlocal vertices.
For example, to Gaussian order
of the expansion of the fermion determinant,
the meson fields can be integrated out and the result defines complex
meson propagators in rainbow-ladder approximation.
{  Masses and widths can be extracted which encode information on
the medium modification of mesons by the underlying
quark-antiquark substructure.}
Therefore, the nl-PNJL, and quark (and gluon) models in general,
are in an interesting position to
properly account for the degrees of freedom in both,
the hadron and the QGP phases, with the underlying physical
mechanism for the vanishing of hadronic states from the spectrum in the QGP
phase being their dissolution in the continuum of scattering states
(the Mott effect)
\cite{Hufner:1994ma,Zhuang:1994dw,Hufner:1996pq,Blaschke:2011hm,
Turko:2011gw,Wergieluk:2012gd,Yamazaki:2012ux,Blaschke:2013zaa}.

In the present work, we are going to develop the nl-PNJL approach further
in three directions.
First, we extend the model with wave function renormalization (WFR)
in a simple way such that it accommodates the medium-induced Lorentz symmetry
breaking (LSB) in the quark propagator invariants.
In order to observe the magnitude of LSB, selected thermodynamic quantities
are displayed together with the scenario which employs only Lorentz symmetric
current-current interactions \cite{Contrera:2010kz,Hell:2011ic}.
More precisely, we compute the quark mean fields at finite temperature
and observe that Lorentz symmetry is heavily broken around and above $T_c$.
In addition, we fill a gap in the literature by providing some analytic
estimates on the effect of WFR and LSB on the critical properties of nonlocal,
as well as local NJL models.

Second, we investigate the role of the PL coupling in this context.
A strong effect of WFR and LSB is to be seen above the chiral pseudocritical
temperature $T_c$.
This leads us to consider the equation of state (EoS) for quark matter, as correlations above could
help maintaining the EoS well below the Stefan-Boltzmann
value even for temperatures up to $0.6-0.8$ GeV as observed in lattice QCD
\cite{Borsanyi:2012rr}.
We use three sets of parametrizations for nonlocality provided in
Ref.~\cite{Contrera:2010kz}.
We demonstrate here for the first time that the behaviour of the EoS is much
more similar to the one measured in lattice QCD simulations in all these cases
only when coupled to the PL.

In our study of LSB we find that even though the
Lorentz covariance of the propagator is drastically broken
above $T_c$,
the bulk thermodynamic properties remain practically untouched.
The critical line in the phase diagram and, especially, the critical
end point (CEP), as well as the EoS, are very little affected by LSB.

Finally we develop our model beyond the mean field by taking into
account Gaussian fluctuations of the pion and sigma mesons.
One novel result is a closed formula for the imaginary part
of the meson polarization loop extracted at zero meson momenta, leading
to the meson width.
By calculating also the meson masses we are able to make an exploratory
study of the Mott effect in a nl-PNJL.
Our results are of technical nature exposing a surprising 
sensitivity to the specific form of the non-local
interactions.
Whereas both the standard non-local interaction, inspired by the separable
DSE model, and the one inspired by the ILM, are equivalent on the mean-field
level, the treatment of fluctuations is somewhat different, 
see e. g. Ref.~\cite{GomezDumm:2006vz}.
We find that this difference leads to dramatically different results for
the meson widths: in the former case the widths start rising, but in the high
temperature regime drop to zero, whereas in the latter case, they 
are monotonous functions of the temperature.

We organize this paper as follows. In Section \ref{sec:model}
the nl-NJL model is shorty reviewed, in order to introduce LSB terms.
Critical properties are discussed in Section~\ref{sec:vac_crit},
notably {the CEP and the phase diagram},
followed by results for the EoS in Section~\ref{sec:eos}.
Beyond the mean field thermodynamics is developed in Section~\ref{sec:mes},
{
whereby details of the mathematical formalism in obtaining the in-medium
mesonic polarization function are separated in an Appendix.
In Section~\ref{sec:concl} we present our Conclusions from the results of
these investigations.}

\section{Setting up the model}
\label{sec:model}

Starting point {of our investigation is the Euclidean action functional
of the nl-NJL} model \cite{Contrera:2010kz}
\be
S_E = \int d^4 x  \left\{\bar{q}(-i\slashchar{\partial}+m)q
-\frac{G_S}{2}\left[j^S_a(x)j^S_a(x)+j_p(x)j_p(x)\right]\right\}~,
\label{nnjl}
\ee
with currents
\be
j^S_a(x)=\int d^4 z g(z)\bar{q}\left(x+\frac{z}{2}\right)
\Gamma_aq\left(x-\frac{z}{2}\right)~,\quad
j_p(x)=\int d^4 z f(z)\bar{q}\left(x+\frac{z}{2}\right)
\frac{i\overleftrightarrow{\slashchar{\partial}}}{2\kappa_p}
q\left(x-\frac{z}{2}\right)~,
\label{crts}
\ee
where $\Gamma_a = (1,i\gamma_5\boldsymbol{\tau})$, and $\boldsymbol{\tau}$ are
Pauli matrices.
When calculating the EoS and the meson properties (see Sections \ref{sec:eos} and \ref{sec:mes}, respectively) 
we will be interested also in a version of the nl-PNJL
inspired by the ILM model.
In this case only the $j_a^S(x)$ current is present in the action (\ref{nnjl}) 
in the form
\be
j^S_a(x)=\int d^4y d^4 z r(y-x)r(x-z)
\bar{q}\left(y\right)\Gamma_aq\left(z\right)~.
\label{crts_ilm}
\ee

We work with $N_f = 2$, $q=(u,d)^\mathrm{T}$.
The symbol $\overleftrightarrow{\partial}_\mu$ provides a shorthand for
$$\psi(x)\overleftrightarrow{\partial}_\mu\phi(y) =
\psi(x)\frac{\partial \phi(y)}{\partial y_\mu}-
\frac{\partial \psi(x)}{\partial x_\mu}\phi(y) ~ .$$
The definite shapes of the regulators $g(z)$, 
$f(z)$ or $r(z)$ in the ILM case, will be provided below in momentum space.
Physically, they can be thought of as mimicking
effective nonlocal 4-quark interactions, or alternatively as wave
functions of quark-antiquark correlations
(see, {\it e.g.}, \cite{Burden:1996nh}).

Finite temperature and chemical potential are introduced via the
Matsubara formalism \cite{Kapusta:1989tk}
analogous to the case of the local NJL model
\cite{Klimt:1989pm,Klevansky:1992qe,Hatsuda:1994pi,Buballa:2003qv}.
The thermodynamic potential in a mean field approximation is
\be
\Omega=\Omega_\mathrm{cond}+\Omega_\mathrm{kin}~,
\label{eq:ommf}
\ee
\ba
\Omega_\mathrm{cond}&=&\frac{1}{2G_S}(\sigma_1^2+\kappa_p^2\sigma_2^2)~,\\
\Omega_\mathrm{kin}&=&- \frac{d_q}{4}
T\sum_{n = -\infty}^\infty \int\frac{d^3p}{(2\pi)^3}
\mathrm{tr}_D\log\left[S^{-1}(\tilde{p}_n)\right]~,
\label{mf1}
\ea
where $\mathrm{tr}_D$ is the Dirac trace, and
$d_q = 2\times 2\times N_c \times N_f$.
The regularization of this divergent quantity is performed as
in \cite{Contrera:2010kz}
providing $\Omega_\mathrm{reg}$.
The full quark propagator is
\be
S^{-1}(\tilde{p}_n) =
-(\gamma\cdot\tilde{p}_n)~ A(\tilde{p}_n^2)+B(\tilde{p}_n^2)~,
\label{eq:qprls}
\ee
where
$\tilde{p}_n^2 = \mathbf{p}^2+\tilde{\omega}_n^2$,
$\tilde{\omega}_n=\omega_n-i\mu$, $\omega_n = (2n+1)\pi T$,
with dressing functions
\ba
A(p^2) &=& 1+\sigma_2 f(p^2)~, \\
B(p^2) &=& m+\sigma_1 g(p^2)~,
\label{symm}
\ea
encoding effect of the background fields $(\sigma_1, \sigma_2)$.
 
For the ILM, the thermodynamic potential on the mean-field level 
takes the same form, provided
that only the scalar channel is kept, i. e. $A(p^2)=1$, and 
a replacement $g(p^2)\to r^2(p^2)$ is
performed.

This kind of quark propagator is very typical for DSE
studies as, {\it e.g.}, in Ref.~\cite{Roberts:2000aa}.
Closest analogy is provided using the separable kernel for the gluon
propagator, as in \cite{Plant:1997jr,Blaschke:1999ab,Blaschke:2000gd}.
Then, one can start from the rainbow-ladder approximation
(RLA) \cite{Roberts:2000aa} of the Cornwall-Jackiw-Tomboulis (CJT)
2PI effective action
\cite{Cornwall:1974vz} of the quark sector and introduce a separable
gluon propagator in order to obtain an expression \cite{Horvatic:2010md}
constructively very similar to (\ref{eq:ommf}).

The regulators specified in
\cite{Noguera:2008cm,Contrera:2010kz}
are dubbed set A (Gaussian, without WFR),
set B (Gaussian, with WFR), and set C
(Lorentzian, with WFR)  {as described below}.
As a shorthand, we also adopt the terminology of separable models as used in
\cite{Blaschke:2000gd,Horvatic:2010md},
referring to models without WFR as rank-1, and to those with WFR as rank-2.
The three regulator sets are defined as
\begin{eqnarray}
\left.
\begin{array}{l}
g(p^2)= \mbox{exp}\left(-p^2/\Lambda_{0}^{2}\right) \\
f(p^2)=0
\end{array}
\right\} \quad {\rm (Set\, A)}~,
\label{eq:setA}
\end{eqnarray}
\begin{eqnarray}
\left.
\begin{array}{l}
g(p^2)= \mbox{exp}\left(-p^{2}/\Lambda_{0}^{2}\right) \\
f(p^2)= \mbox{exp}\left(-p^{2}/\Lambda_{1}^{2}\right)
\end{array}
\right\} \quad {\rm (Set\, B)}~,
\label{eq:setB}
\end{eqnarray}
\begin{eqnarray}
\left.
\begin{array}{l}
g(p^2)  = \frac{1+\alpha_z}{1+\alpha_z\ f_z(p^2)}
\frac{\alpha_m \ f_m (p^2) -m\ \alpha_z f_z(p^2)}
{\alpha_m - m \ \alpha_z } \\
f(p^2)  = \frac{ 1+ \alpha_z}{1+\alpha_z \ f_z(p^2)} f_z(p^2)\
\end{array}
\right\} \quad {\mathrm{ (Set\, C)}}~,
\label{eq:setC}
\end{eqnarray}
where
\ba
f_{m}(p^2) &=& \left[ 1+ \left( p^{2}/\Lambda_{0}^{2}\right)^{3/2} \right]^{-1}
~, \\
f_{z}(p^2) &=& \left[ 1+ p^{2}/\Lambda_{1}^{2} \right]^{-5/2}~,
\label{eq:setC2}
\ea
and $\alpha_m = 309$ MeV, $\alpha_{z}=-0.3$.
For ILM model, we have
\begin{eqnarray}
\left.
\begin{array}{l}
r(p^2)= \mbox{exp}\left(-p^2/2\Lambda_{0}^{2}\right) \\
f(p^2)=0
\end{array}
\right\} \quad {\rm (ILM)}~.
\label{eq:ILM}
\end{eqnarray}
All the parameter sets are summarized in Table~\ref{tab:parameters}.
\begin{table}[htb]
\begin{center}
 \begin{tabular}{|c|c|c|c|c|}
   \hline
   -- & Set A & Set B & Set C & ILM \\ \hline
   $m$ [MeV] & 5.78 & 5.7 & 2.37  & 5.8\\
   $\Lambda_0$ [GeV] & 0.752 & 0.814 & 0.850 & 0.902\\
   $G_S\,\Lambda_0^2$ &  $20.65$ & $32.03$ & $20.818$ & $15.82$ \\
   $\Lambda_1$ [GeV] & -- & 1.034 & 1.400.0 & --\\
   $\kappa_p$ [GeV] & -- & 4.180 & 6.034 & --\\
   \hline
 \end{tabular}
\end{center}
 \caption{Parameter sets A -- C, and the ILM model as used in this work.
For further details on set A -- C, 
see Refs.~\cite{Noguera:2008cm,Contrera:2010kz}, and for the ILM model
see \cite{GomezDumm:2006vz}.}
\label{tab:parameters}
\end{table}

\subsection{Lorentz symmetry breaking by the medium}
\label{lvsec}

As the medium presents a distinct
reference frame, Lorentz symmetry is broken.
Effects of this breaking are revealed in the richer
tensor structures
for the Green's functions of the theory,
notably the propagators.
Here we explore the possibility of splitting the
WFR term in the quark propagator.
This is a very well known effect in DSE studies at finite temperatures and
chemical potentials \cite{Bender:1996bm,Blaschke:1998md}, see also
\cite{Roberts:2000aa}, through which, for example,
the possible existence of plasmino modes
above $T_c$ can be explored \cite{Mueller:2010ah,Qin:2010pc}.

The residual $O(3)$ symmetry of the medium allows the following structure
of the quark propagator\footnote{Here we have two vectors at our
disposal: the momentum of the particle, and the momentum of the
medium.
Therefore, there may be in principle medium-induced
tensor forces (see e. g. \cite{Kalashnikov:1980bm}) giving rise to
a $\sigma^{\mu\nu}$ term in the propagator.
To get this term one should include a tensor channel in the NJL
model, a possibility which we do not consider in this work.}
\be
S^{-1}(\tilde{p}_n)= -(\boldsymbol{\gamma}\cdot\mathbf{p})~ A(\tilde{p}_n^2)
-\gamma_4 \tilde{\omega}_n C(\tilde{p}_n^2)+B(\tilde{p}_n^2)~.
\label{qprop2}
\ee
It is clear that a covariant nl-NJL model interaction $j_p(x) j_p(x)$
see (\ref{nnjl}), can only yield $C(p^2) = A(p^2)$ (\ref{symm}).
In order to take into account also the more general possibility $A(p^2)\neq C(p^2)$,
we break the $O(4)$ symmetry to $O(3)$ in the
interaction itself by modifying the $j_p(x)$ channel
\be
j_p j_p \to j_\mathbf{p}j_\mathbf{p} + j_{p_4} j_{p_4}~,
\ee
where
\ba
j_\mathbf{p}(x)&=&\int d^4 z f(z)\bar{q}\left(x+\frac{z}{2}\right)
\frac{i\overleftrightarrow{\nabla}\boldsymbol{\gamma}}{2\kappa_\mathbf{p}}
q\left(x-\frac{z}{2}\right)~,\\
j_{p_4}(x)&=&\int d^4 z f(z)\bar{q}\left(x+\frac{z}{2}\right)
\frac{i\overleftrightarrow{\partial_4}\gamma_4}{2\kappa_{p_4}}
q\left(x-\frac{z}{2}\right)~,
\ea
with the couplings $\kappa_\mathbf{p}$ and $\kappa_{p_4}$ regulating
the strength of each term.
This modification now preserves only $O(3)$ symmetry, and
alters the thermodynamic potential (\ref{mf1}).
The condensate term $\Omega_\mathrm{cond}$ becomes
\be
\Omega_\mathrm{cond} \to
\frac{1}{2G_S}\left(\sigma_B^2 + \kappa_\mathbf{p}^2 \sigma_A^2
 + \kappa_{p_4}^2 \sigma_C^2 \right)~,
\label{cdnew}
\ee
while the quark propagator in $\Omega_\mathrm{kin}$ goes to (\ref{qprop2}).
In discussing the effects of LSB we use for the mean fields the same nomenclature as in Ref.\cite{Roberts:2000aa}, i.e. $\sigma_i$ $(i=A,B,C)$, in order to differentiate from $\sigma_{1,2}$ of the LS case.
$C(p^2)$ is yet another quark dressing function symbolizing breakdown of $O(4)$
symmetry $C(p^2) = 1+\sigma_C f(p^2)$.

Full correspondence with the separable DSE studies in, {\it e.g.}
Refs.~\cite{Blaschke:2000gd,Horvatic:2010md}, is obtained by using
$\kappa_\mathbf{p}^2/\kappa_{p_4}^2=3$.
In order to restore the $O(4)$ symmetric form (\ref{mf1}) in 
the vacuum we must have
$\kappa_\mathbf{p}^2 = 3\kappa_p^2/4$, $\kappa_{p_4}^2 = \kappa_p^2/4$.

\subsection{Polyakov loop}

The PL \cite{Polyakov:1978vu} $\Phi$ (and it's conjugate $\bar{\Phi}$)
represents a non-perturbative pure-glue vacuum response to
an infinitely heavy ``probe'' quark (antiquark).
As such, it stands for an
order parameter for confinement in accordance with the spontaneous
breaking of center symmetry of the gauge group $SU(3)_c$.
However, the center symmetry is strictly broken with dynamical quarks winding
around the thermal circle as they are bound to respect the
antiperiodic boundary conditions.

The PL is introduced as the color trace over a position independent timelike
gluon background field $\phi_3$ in the Polyakov gauge \cite{Polyakov:1978vu},
$\Phi = [1+2\cos(\phi_3/T)]/N_c$,
which modifies the Matsubara frequencies
$\tilde{\omega}_n = \omega_n - i\mu + \lambda_3 \phi_3$,
depending on the color state.
In the thermodynamic potential the color trace, as well as
the Dirac trace, becomes non-trivial
\be
\Omega_\mathrm{kin} =
-\frac{d_q}{12} T\sum_{n=-\infty}^\infty\int\frac{d^3 p }{(2\pi)^3}
\mathrm{tr}_{D,C}\log S^{-1}( \tilde{p}_n) ~.
\label{kin2}
\ee
The unregularized mean field 
thermodynamic potential is then augmented
by a gluon mean field potential {$\mathcal{U}(\Phi,T)$ } to become
\be
\Omega = \Omega_\mathrm{cond}+\Omega_\mathrm{kin}
+\mathcal{U}(\Phi,T)~,
\label{ommf}
\ee
where we choose the logarithmic form of the PL potential
{$\mathcal{U}(\Phi,T)$ } introduced in \cite{Roessner:2006xn}

{
\begin{equation}
\mathcal{U}(\Phi,T) = \left[-\,\frac{1}{2}\, a(T)\,\Phi^2 \;+\;b(T)\, \ln(1
- 6\, \Phi^2 + 8\, \Phi^3 - 3\, \Phi^4)\right] T^4 \ ,
\end{equation}
with
$a(T)=a_0+a_1(T_0/T)+a_2(T_0/T)^2$~,
$b(T)=b_3(T_0/T)^3$~.
The corresponding parameters are $a_0=3.51$, $a_1=-2.47$, $a_2=15.22$ and $b_3=-1.75$.
In the present work we set $T_0= 0.27$ GeV.
}

\subsection{Physical meaning of the mean fields}

The $\sigma_B$ mean field is closely related
to the quark-condensate $\langle \bar{q} q\rangle$
signalling chiral symmetry breaking.
Although in the nl-NJL the mass is a dynamical
quantity, depending on quark momentum, $\sigma_B$ is usually
referred to as the mass gap.

The ``derivative'' mean fields, $\sigma_A$
and $\sigma_C$, provide the quark propagator
with a nonzero WFR as seen
on the lattice as well as in DSE models.
It is very useful to consider the NJL-like limit
of the model with  $f(p^2)\to \theta(\Lambda_0^2-\mathbf{p}^2)$ and $g(p^2) \to \theta(\Lambda_0^2-\mathbf{p}^2)$.
The NJL thermodynamic potential with WFR and LSB
can be simply obtained from the one without the WFR given in,
{\it e.g.}, Ref.~\cite{Buballa:2003qv}.
While $\Omega_\mathrm{cond}$ can be directly
taken from Eq.~(\ref{cdnew}),
the kinetic part is the  quasiparticle Fermi gas,
\be
\Omega_\mathrm{kin} = - \frac{d_q}{2}\int\frac{d^3 p}{(2\pi)^3}
\left\{E + T\log[1+e^{-\beta(E-\mu)}]+T\log[1+e^{-\beta(E+\mu)}]\right\}~,
\label{njl_wfr}
\ee
where $E$ is given by
\be
v_\mathrm{qp}^2\mathbf{p}^2 - E^2 + m_\mathrm{qp}^2 = 0~,
\label{quasi_disp}
\ee
and
\be
v_\mathrm{qp} = \frac{A_0}{C_0} = \frac{1+\sigma_A}{1+\sigma_C} \, , \qquad
m_\mathrm{qp} = \frac{B_0}{C_0} = \frac{m+\sigma_B}{1+\sigma_C}~,
\label{disp}
\ee
where $A_0=A(0)$, $B_0=B(0)$ and $C_0=C(0)$.
The values $1/A_0$ and $1/C_0$ represent WFR.
Furthermore, causality requires $v_\mathrm{qp}\leq 1$ (the speed of light)
leading to $\sigma_A \leq \sigma_C$.
This is the first physical manifestation of LSB
encoded in the full numerical solutions
in the following sections.

The most important use of the PL in NJL models is to suppress quark excitations
at low temperatures.
In covariant nl-NJL models remnants of the quark excitations are still present
in the complex plane in the confining phase, leading to unphysical
thermodynamic behavior.
The PL then acts to strongly suppress such states
from being thermally excited \cite{Benic:2012ec}, 
see also subsection~\ref{sub_instab}.

\section{Critical properties}
\label{sec:vac_crit}

In this section we discuss the effect of the wave function
renormalization on the critical coupling for chiral symmetry breaking and the
chiral restoration temperature.
We restrict ourselves to discuss only sets A -- C, the ILM model
will become important in the following Sections.
The following analytical estimates are restricted to the chiral limit and to
the case without the PL.
Next, solutions of the gap equations with and without LSB effects will be
compared.
Results show that LSB is more profound around the chiral restoration, in
accordance with \cite{Blaschke:2000gd}.
Finally, the influence of LSB on the phase diagram and on the CEP is
calculated.

\subsection{Critical coupling analysis}

In this subsection we work in the chiral limit $m=0$.
The onset of the chiral transition is controlled by the strength of the
scalar channel $G_S$.
For the local NJL with a standard 3D cutoff $\Lambda_0$,
the critical value for the coupling is \cite{Yagi:2005yb}
\be
G_S^c \Lambda_0^2 = \frac{8\pi^2}{d_q}~.
\label{critnjl}
\ee
One can easily show that the effect of a constant
WFR
amounts to
\be
G_S^c \Lambda_0^2 = \frac{8\pi^2}{d_q}A_0^2~,
\label{critnjl_wfr}
\ee
where the term $A_0=1+\sigma_2>1$ leads to an
increase in the critical coupling.

In rank-1 Gaussian models we quote \cite{GomezDumm:2004sr}
the following result $G_S^c \Lambda_0^2 = 4 \times 8\pi^2/d_q$,
while for rank-2 Gaussian models (set B) one can
obtain a similar expression
\be
G_S^c \Lambda_0^2 = 4~\frac{8\pi^2}{d_q}
\frac{1}{\rho\left(\sigma_2,\frac{\Lambda_0^2}{\Lambda_1^2}\right)}~,
\label{gcritr2}
\ee
where
\be
\rho(a,x) = 2\int_0^\infty\ dy \ \frac{ye^{-y^2}}{(1+ae^{-xy^2/2})^2} =
 1-\frac{2a}{1+\frac{x}{2}}+\frac{3 a^2}{1+x}+\dots ~,
\label{ar}
\ee
and $\Lambda_0$ and $\Lambda_1$ are the scales of the appropriate
regulators, see Eqs.~(\ref{eq:setA})-(\ref{eq:ILM}).
The second equality provides an expansion in $\sigma_2$, valid
for $\sigma_2 <1$.
Then $\rho<1$, and we have
$(G_S^c \Lambda_0^2)_\mathrm{rank-2}>
(G_S^c \Lambda_0^2)_\mathrm{rank-1}$,
concluding that the critical coupling is in
principle always larger
for rank-2 than for rank-1.
This is in accord with the above simplified NJL scenario.
If we are to use some reasonable values, say
$\Lambda_0 \simeq \Lambda_1$ and
$\sigma_2 \sim 0.5$, we have $\rho \simeq 0.6$.

\subsection{Critical line in the phase diagram}

Let us now proceed to approximate the influence of wave function
renormalization on $T_c$.
In the local \cite{Yagi:2005yb},
as well as rank-1 nl-NJL \cite{GomezDumm:2004sr}, this is simply given as
\be
T_c = \left[\frac{24}{d_q}\left(\frac{1}{G_S^c}-
\frac{1}{G_S}\right)\right]^{1/2} ~,
\ee
with $G_S^c$ given by their respective values.

With WFR, the analysis is very similar.
The quark loop that needs to be evaluated is
\be
\frac{\partial^2\Omega_\mathrm{kin}}{\partial\sigma_B^2}\Big|_{\sigma_B = 0} =
-d_q T \sum_{n=-\infty}^\infty \int\frac{d^3 p}{(2\pi)^3}
\frac{g^2(p_n^2)}{\mathbf{p}^2A^2(p_n^2)+\omega_n^2 C^2(p_n^2)}~.
\label{loop}
\ee
We first study a slightly simplified scenario with $A=C$.
Then, the denominator as a function of $z=i\omega_n$
has simple poles at $\pm \mathbf{p}$.
For rank-1, these are the only poles.
For rank-2 set B, there is also an infinite
tower of double poles
when $A^2(-z^2 + \mathbf{p}^2) = 0$.
We ignore them at this point, by assuming they have a significant effect
only after $T_c$, see Sec. \ref{sec:eos}.
By explicitly evaluating the Matsubara sum, as well as the momenta
integral, we obtain
\be
T_c \simeq A_0\left[\frac{24}{d_q}\left(\frac{1}{G_S^c}-\frac{1}{G_S}\right)\right]^{1/2}~,
\ee
with $G_S^c$ given by (\ref{gcritr2}).

If we suppose that $G_S$ in rank-2 is scaled to $G_S$ in rank-1,
just like it is true for
$G_S^c$ (see Eq. (\ref{gcritr2})), and that also, for simplicity,
cutoff scales $\Lambda_0$ are the same,
we conclude that
$$(T_c)_\mathrm{rank-2} \simeq
A_0 \, \rho^{1/2}\left(\sigma_2,\frac{\Lambda_0^2}{\Lambda_1^2}\right) \, (T_c)_\mathrm{rank-1}~,$$
where $\rho(a,x)$ is given in (\ref{ar}).
As $\rho<1$ and $A_0>1$, there occurs a compensation, producing
roughly the
same temperature as in rank-1.
Taking actual values for set B
\cite{Noguera:2008cm,Contrera:2010kz},
we obtain $(T_c)_\mathrm{rank-2} \simeq 1.08 (T_c)_\mathrm{rank-1}$.

The first non-trivial effect of LSB on $T_c$ can be established
by studying (\ref{loop}) for poles $\pm v_\mathrm{qp}\mathbf{p}$.
It is an easy task to show that
\be
(T_c)_\mathrm{rank-2}^\mathrm{LSB} \simeq
v^{1/2}_\mathrm{qp} \, (T_c)_\mathrm{rank-2}^\mathrm{LS}~,
\label{tcrt}
\ee
where $(T_c)_\mathrm{rank-2}^\mathrm{LS}$ is provided by the
previous equation.
Therefore, LSB leads to a decrease of the critical temperature.

Introducing the chemical potential can lead to a change in the critical
behavior - from second order at low $\mu$ to a first order transition
at high $\mu$.
We ask for a simplest possible analytic estimate on
the effect of WFR and LSB on
the phase transition line and on the CEP.
Therefore, we will show explicit analytic results only in the local NJL limit.
A high temperature expansion
\cite{Actor:1986zf}, of (\ref{njl_wfr})
leads to a Landau form of the thermodynamic
potential, i. e.
\be
\Omega \simeq -\frac{1}{2} D(T,\mu)\sigma_B^2 + \frac{1}{4}F(T,\mu)\sigma_B^4~.
\ee
The Landau coefficients, $D(T,\mu)$ and $F(T,\mu)$, are
\be
D(T,\mu) = -\frac{1}{G_S}+\frac{1}{v_\mathrm{qp}^3}\frac{1}{C_0^2}
\frac{1}{G_S^c}+
\frac{d_q}{8\pi^2}\frac{T^2}{v_\mathrm{qp}^3}
\left(\frac{\pi^2}{3}+\frac{\mu^2}{T^2}\right)
\frac{1}{C_0^2}~,
\label{land1}
\ee
\be
F(T,\mu) \simeq
\frac{d_q}{8\pi^2}\frac{1}{v_\mathrm{qp}^3}\left[
\log\frac{\Lambda_0}{2\pi T}+\gamma-1+\frac{7}{2}\zeta(3)
\left(\frac{\mu}{2\pi T}\right)^2\right]
\frac{1}{C_0^4}~,
\label{land2}
\ee
with $G_S^c$ given by (\ref{critnjl}).
We should warn that, while (\ref{land1}) is exact, in (\ref{land2})
we restrict ourselves only to the first
non-trivial term in the $\mu/T$ expansion \cite{Actor:1986zf}.

Requiring $D(T,\mu)=0$ gives us the behavior of the critical line
$T_c(\mu)$ at $\mu/T \ll 1$.
A canonical form is established by
\be
\frac{T_c(\mu)}{T_c(0)} =
1-\kappa \left(\frac{\mu}{T_c(\mu)}\right)^2~,
\ee
where $\kappa$ denotes the curvature of the critical line.
The importance of this quantity lies in the fact
that it can be measured on the lattice,
see e. g. \cite{Kaczmarek:2011zz}.
We can immediately see that introducing WFR, as well as LSB
does not change the curvature,
the latter being simply $\kappa = 3/\pi^2$.
The same can be conjectured also for the nonlocal rank-1 models,
because the medium
component of (\ref{land1}) is governed by the singularities
of the quark propagator.
At $\sigma_1 = 0$, the nonlocal rank-1 model has the same
singularities as the local one.
However, as was already mentioned, rank-2 models have additional
singularities in the WFR term,
see (\ref{loop}), which might then alter the medium part in (\ref{land1}).
In fact, a full numerical study \cite{Contrera:2012wj} shows
that the general effect of WFR is to
increase $\kappa$.
From the physical point of view this is to be expected,
since the singularities effectively
act as additional ``degrees of freedom''.

The CEP
can be inferred by simultaneously requiring $D(T,\mu)=0$ and $F(T,\mu)=0$.
Restricting to keep only $\sim(\mu/T)^2$ term in (\ref{land2}) limits
the discussion somewhat
by excluding a possible low $T$, high $\mu$ CEP.
On the other hand, by inserting (\ref{land1}) into (\ref{land2})
we are lead to a
simple condition on $T_\mathrm{CEP}$
\be
-\log \frac{T_\mathrm{CEP}}{\Lambda_0} +
\frac{7\zeta(3)}{24}\frac{T_c^2(0)}{T_\mathrm{CEP}^2} =
\frac{7\zeta(3)}{24}+\log(2\pi)+1-\gamma \equiv R~,
\ee
which can be easily analyzed.
Now we may estimate the influence of WFR and LSB on the CEP.
First of all, the right hand side of the last equation is
a pure number, $R\simeq 2.61$.
Second, since we know that $T_c(0) \simeq v_\mathrm{qp}^{1/2}
T^{\mathrm{LS}}_c(0)$, the quadratically divergent
term will be somewhat stronger,
further decreasing $T_\mathrm{CEP}$.
If we take this term to be the dominant one, we obtain
\be
T_\mathrm{CEP}^{\mathrm{LSB}} \simeq
v_\mathrm{qp}^{1/2}T_\mathrm{CEP}^{\mathrm{LS}}=
\sqrt{\frac{7 \zeta(3)}{24 R}} \, v_\mathrm{qp}^{1/2} \, T_c^{\mathrm{LS}}(0) \simeq
0.37 \, v_\mathrm{qp}^{1/2} \, T_c^{\mathrm{LS}}(0) ~.
\ee
{  Within this approximation, there is no
influence of the LS version of the WFR channel on the CEP.
On the other hand, we may conclude that the first
estimate on the influence of
the LSB on the CEP is that the CEP
goes to lower $T$ and, consequently, to higher $\mu$.}
Owing to the fact that the presented analysis is rather crude,
and formally confined to $\mu/T\ll 1$,
we conclude that deviations might be even larger, and get further increased
in the rank-2 nonlocal case.
A full numerical study in nonlocal models, see
e. g. \cite{Contrera:2010kz,Contrera:2012wj}, supports this conjecture.

\subsection{Splitting of $\sigma_A$ and $\sigma_C$ and the phase diagram}

\begin{figure}[t]
\begin{center}
\parbox{15cm}{
\psfig{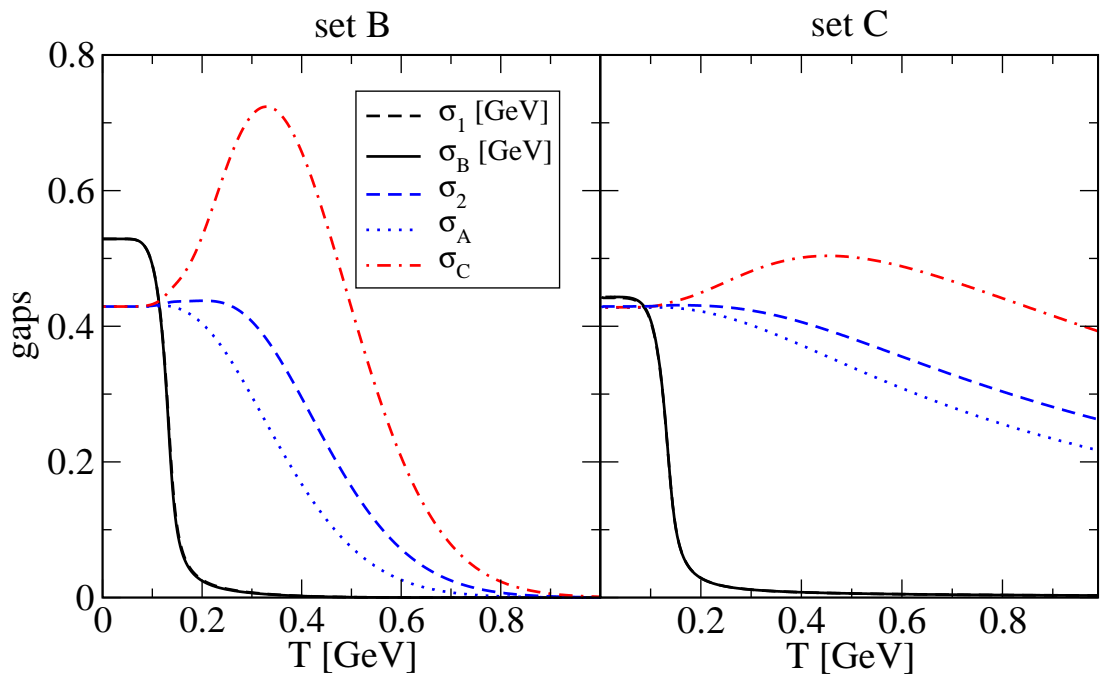}
}
\caption{(Color online) We illustrate the effect of LSB in set B
(left) and set C (right) at $\mu=0$.
For simplicity, the system has been solved without PL.}
\label{lorentzac}
\end{center}
\end{figure}
In Section \ref{lvsec} we argued for the possibility of the most
general structure of the quark propagator (\ref{qprop2}).
The numerical results obtained from the minimization of the
thermodynamic potential are shown in Fig. \ref{lorentzac}.
By comparing the mass gaps, it is plain  that there is barely
an influence.

On the contrary, in Fig. \ref{lorentzac}, there is a clear difference
between $\sigma_A$ and $\sigma_C$ mean fields
defining a region where $O(4)$ symmetry is violated.
This difference is a
reflection of the $R^3 \times S^1$ structure of the spacetime manifold,
and was already observed in DSE separable model studies, e. g.
\cite{Blaschke:2000gd,Horvatic:2007wu}.
At low temperatures the thermal circle $S^1$ is large, and Lorentz symmetry
is approximately valid.
With the increase in the temperature, $\sigma_A$ and
$\sigma_C$ split, the difference is starting to be
pronounced around the phase
transition as the gap equations form
a coupled system.
Namely, since around the phase transition the mass gap suffers a
significant drop, this
must be reflected in changes of the gaps $\sigma_A$ and $\sigma_C$.
We see that the particular behavior of the mean fields is
``causal'', governing the inequality $\sigma_A<\sigma_C$.

From Fig. \ref{lorentzac} we conclude that the splitting
is much stronger for set B; in the region
$0.2 \, \mathrm{GeV} \lesssim T\lesssim 0.6 \, \mathrm{GeV}$
$\sigma_C$ develops a pronounced peak, whereas $\sigma_A$
monotonously descends.
The value of $\sigma_2$ in the LS case can then be understood to provide
a ``mean value'' between these two behaviors.
The most distinct characteristic of the mean fields in set C
is the finite value of $\sigma_A$ and $\sigma_C$, referring
to highly non-perturbative quarks even at $T\approx 1$ GeV!

The phase diagrams in this model for sets A, B, C were
presented in \cite{Contrera:2010kz}.
We are interested in the effect of the splitting of $\sigma_A-\sigma_C$
on the phase transition line, most notably on the position of the CEP.

The order parameter of chiral symmetry breaking
is the quark condensate
\be
\langle \bar{q}q\rangle =
\frac{\partial \Omega_\mathrm{reg}}{\partial m}~.
\ee
The pseudocritical temperature $T_\mathrm{c}$ in the crossover transition
region is conveniently defined as in \cite{Contrera:2010kz},
with the temperature where the chiral susceptibility
$\chi = \partial \langle \bar{q}q \rangle/\partial m$
is maximal.
For the first order region, the point where the
chirally broken and chirally restored solution of the gap equation
have the same value of the thermodynamic potential,
defines the transition point in the phase diagram.
This way, a curve $T_c(\mu)$ in the $T-\mu$ plane is provided.

Even though the mass gap is practically identical in both setups,
see Fig. \ref{lorentzac},
quark condensate is also affected by $\sigma_A$ and $\sigma_C$,
thereby some difference in the critical line is to be anticipated.
However, we do not expect the actual change to be drastic,
as the condensate
is mostly driven by the value of the mass gap.

\begin{figure}[!t]
\begin{center}
\parbox{14cm}{
\psfig{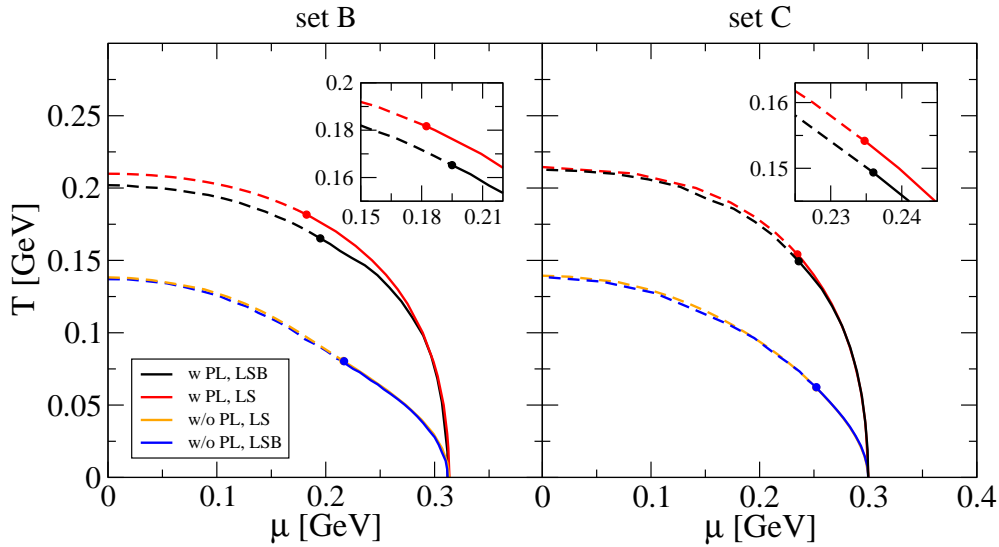}
}
\caption{(Color online) Left (right) panel:
phase diagrams for set B (set C)  for cases with and without PL.
With PL and LSB the results are shown in black, while the results from
Ref. \cite{Contrera:2010kz} are repeated
in red.
Blue (orange) lines are results without PL, and with (without) LSB.
Dashed line denotes crossover, and the full line is the
first order transition.}
\label{phasediag}
\end{center}
\end{figure}

Fig. \ref{phasediag} shows results for the phase diagrams
of rank-2 models: set B and set C in both cases.
Some general remarks are in order.
First, the presence of the PL increases the pseudocritical
temperature $T_c(0)$ in both models by $\sim 50$ MeV.
This can be argued by a simple analytical
formula provided by \cite{Contrera:2010kz}, and from the fact
that the pure YM sector provides a transition temperature of
$T_0 = 0.27$ GeV \cite{Roessner:2006xn}.
Second, the first order transition of the pure Yang-Mills (YM)
sector ``pushes''
the CEP closer to the $T$ axes.
Finally, the effect of the PL is less significant once the temperature
is sufficiently low.
Critical lines of both cases, with and without PL,
join at $T=0$.

The explicit value of $T_c(0)$ is somewhat high, being around $0.2$ GeV
regardless of the model details, while lattice results
for two flavors \cite{Ejiri:2000bw} provide a value of 0.17 GeV.
This can be easily amended by rescaling the YM critical temperature $T_0$ as
argued in \cite{Schaefer:2007pw}.
Within nl-PNJL models, the effect of such a rescaling on the
pseudocritical temperature $T_c(0)$ and on the width of the transition
has been explored, {\it e.g.}, in  \cite{Horvatic:2010md,Radzhabov:2010dd}.
In Ref. \cite{Contrera:2012wj} an account on the
phase diagram in nl-PNJL, with rescaled
$T_0$ can be found (see also \cite{Hell:2012da}).

For set B, the critical lines, as given on left panel
of Fig. \ref{phasediag}, are changed only in
the high $T$, low $\mu$ region.
Specifically,  we obtain a somewhat lower $T_c(0)$ for the LSB case,
in accordance with the analytical estimate (\ref{tcrt}).
Region around CEP is slightly altered, shifting the value of the CEP to
lower $T$ and higher $\mu$ for $\sim 20$ MeV.
We regard the critical lines for set C, on the right panel of
Fig. \ref{phasediag},
as almost identical,
with the LSB curve being only a few MeV below the one reported
in \cite{Contrera:2010kz}.
This is just a reflection of the results in the previous
section, where, at least
for $\mu=0$, Fig. \ref{lorentzac} explicitly shows
that the $\sigma_A-\sigma_C$
splitting is much stronger for set B than for set C.

\section{Finite temperature mean field equation of state}
\label{sec:eos}

In the following section first a brief
summary of results 
found in Ref.~\cite{Benic:2012ec} is highlighted 
in order to 
explain why thermodynamic instabilities 
are in general expected when one deals with covariant quark models.
Technical steps are omitted for brevity.
Moreover, we upgrade the study of the analytic structure of rank-1
models with a Gaussian regulator \cite{Bowler:1994ir,Plant:1997jr,GomezDumm:2001fz}, to rank-2, revealing
a crucial difference between these two models, needed for
understanding the thermal behavior of the EoS.
Finally, we analyze the difference of the EoS with and without LS.

\subsection{Instability in covariant chiral quark models}
\label{sub_instab}

The central quantity is the kinetic contribution to the 
thermodynamic potential
(\ref{mf1}).
In order to understand the principle mechanism it is sufficient to conjecture
that the quark propagator has a series of $P$ 
simple complex conjugate mass poles (CCMPs).
By standard residue analysis \cite{Benic:2012ec} in the case without the PL
one is then able to obtain
\be
\begin{split}
\Omega_\mathrm{kin} &= \Omega_\mathrm{zpt}-4T N_f N_c \sum_{k=1}^P 
\int\frac{d^3 p}{(2\pi)^3}\left[\log(1+e^{-\beta \mathcal{E}_k})+
\mathrm{log}(1+e^{-\beta \mathcal{E}_k^*})\right]\\
&=\Omega_\mathrm{zpt}-4T N_f N_c \sum_{k=1}^P 
\int\frac{d^3 p}{(2\pi)^3}\log[1+
2\cos(\beta\gamma_k)e^{-\beta \epsilon_k}+e^{-2\beta \epsilon_k}]~,
\end{split}
\label{eq:osc}
\ee 
where the 
notation 
$\mathcal{E}_k(\mathbf{p}) = \epsilon_k(\mathbf{p}) + i\gamma_k(\mathbf{p})$ for 
the CCMPs was used.
They are given as
\be
\epsilon_k(\mathbf{p}) = \frac{1}{\sqrt{2}}
\left\{(m^R_k)^2-(m^I_k)^2+\mathbf{p}^2+
\sqrt{\left[(m^R_k)^2-(m^I_k)^2+\mathbf{p}^2\right]^2
+4(m^R_k)^2(m^I_k)^2}\right\}^{1/2}~, 
\label{eq:ccre}
\ee
and
\be
\gamma_k(\mathbf{p}) = \frac{m^R_k m^I_k}{\epsilon_k} = \frac{1}{\sqrt{2}}
\left\{-(m^R_k)^2+(m^I_k)^2-\mathbf{p}^2+
\sqrt{\left[(m^R_k)^2-(m^I_k)^2+\mathbf{p}^2\right]^2
+4(m^R_k)^2(m^I_k)^2}\right\}^{1/2}~, 
\label{eq:ccim}
\ee
where $m_k^R$ and $m_k^I$ are real and imaginary parts of 
complex masses, respectively.
In general, they are functions of the mean fields
\be
m_k^R=m_k^R(\sigma_A,\sigma_B,\sigma_C) \, , \qquad
m_k^I=m_k^I(\sigma_A,\sigma_B,\sigma_C)~.
\ee
The quantity $\Omega_\mathrm{zpt}$ represents the zero-point energy.
With the combined logarithms in the second equality 
it is easily observed that a non-zero value of at least
one $\gamma_k$ leads to an oscillating EoS.
Namely, if the oscillations are expected in the confining, low $T$ domain
one can perform an expansion in $m_k^R/T \gg 1$ of thermal part in (\ref{eq:osc}).
If, in addition one assumes that $m_k^I \ll m_k^R$, then
\be
\Omega_\mathrm{kin} \simeq \Omega_\mathrm{zpt}-4 N_f N_c T^4 \sum_{k=1}^P\left[
2\cos\left(\frac{m_k^I}{T}\right)\left(\frac{m_k^R}{2\pi T}\right)^{3/2}
e^{-m_k^R/T}+\left(\frac{m_k^R}{4\pi T}\right)^{3/2}e^{-2m_k^R/T}\right]~,
\ee
which is a generalization of the low temperature 
expansion \cite{Kapusta:1989tk} for complex masses.

Including the effect of the PL, i. e. performing a Matsubara sum 
in Eq.~(\ref{kin2}), gives
\be
\begin{split}
\Omega_{\mathrm{kin}} &= 
\Omega_\mathrm{zpt}-4 N_f T \sum_{k=1}^{P} \int \frac{d^3 p}{(2\pi)^3} 
\log \Big\{1+6\Phi\big[(e^{-\beta\epsilon_k}+e^{-5\beta\epsilon_k})
\cos(\beta\gamma_k)\\
&+(e^{-2\beta\epsilon_k}+e^{-4\beta\epsilon_k})\cos(2\beta\gamma_k)\big]\\
&+9\Phi^2[e^{-2\beta\epsilon_k}+e^{-4\beta\epsilon_k}+
2e^{-2\beta\epsilon_k}\cos(\beta\gamma_k)]
+2e^{-3\beta\epsilon_k}\cos(3\beta\gamma_k)+e^{-6\beta\epsilon_k}\Big\}~.
\end{split}
\label{pkinpl2}
\ee
reflecting the stabilization mechanism by the PL: in the confining
phase $\Phi \approx 0$ and the oscillating terms are significantly suppressed.
This can be explicitly seen in the low $T$ expansion of (\ref{pkinpl2})
\be
\begin{split}
\Omega_{\mathrm{kin}} &\simeq 
\Omega_\mathrm{zpt}-4 N_f T^4 \sum_{k=1}^{P}
\Bigg\{
6\Phi \cos\left(\frac{m_k^I}{T}\right)\left[\left(\frac{m_k^R}{2\pi T}\right)^{3/2}
e^{-m_k^R/T}+
\left(\frac{m_k^R}{10\pi T}\right)^{3/2}e^{-5m_k^R/T}\right]\\
&+6\Phi \cos\left(\frac{2m_k^I}{T}\right)\left[\left(\frac{m_k^R}{4\pi T}\right)^{3/2}
e^{-2m_k^R/T}+
\left(\frac{m_k^R}{8\pi T}\right)^{3/2}e^{-4m_k^R/T}\right]\\
&+9\Phi^2\left[\left(\frac{m_k^R}{4\pi T}\right)^{3/2}
e^{-2m_k^R/T}+
\left(\frac{m_k^R}{8\pi T}\right)^{3/2}e^{-4m_k^R/T}+
2\cos\left(\frac{m_k^I}{T}\right)\left(\frac{m_k^R}{4\pi T}\right)^{3/2}
e^{-2m_k^R/T}\right]\\
&+2\cos\left(\frac{3m_k^I}{T}\right)\left(\frac{m_k^R}{6\pi T}\right)^{3/2}e^{-3m_k^R/T}+\left(\frac{m_k^R}{12\pi T}\right)^{3/2}e^{-6m_k^R/T}
\Bigg\}~.
\end{split}
\ee

\subsection{Overcritical vs. undercritical mass gaps}
\label{sub_overc}
In the last subsection we have argued that oscillations may appear in the EoS
if at least one $\gamma_k$ is complex.
Now we will make the preparatory analysis in order to be
able to discuss the question in which temperature region that occurs.

To understand the connection between the oscillations and the mass 
gap $\sigma_1$, one traces singularities as functions of $\sigma_1$.
The salient features will be presented for Gaussian regulators,
and in the chiral limit.
We will also restrict the analysis to the lowest 
lying poles as they carry
all the essential properties in the temperature 
range that is discussed.

For rank-1 Gaussian, a value of $\sigma_1 > \sigma_1^c$, where 
$\sigma_1^c = \Lambda_0/(\sqrt{2e})$ gives only 
complex poles
in the propagator, while $\sigma_1 < \sigma_1^c$ gives 
also a pair of real poles.
In set A, 
the vacuum value is overcritical, i. e. $\sigma_1 >\sigma_1^c$,
thus all the poles are complex, and the oscillations are 
present in $T\lesssim T_c$ region.
More concretely, in the chiral limit, we have
$\sigma_1 = 0.402$ GeV and $\sigma_1^c = 0.322$ GeV.
ILM models usually supports weaker 
interaction strengths, as is e. g. the case
for the specific parameters discussed here, see Table~\ref{tab:parameters}.
This typically leads to 
undercritical gaps, for parameters given in Table~\ref{tab:parameters} in the chiral limit 
we have $\sigma_1 = 0.215$ GeV, and 
$\sigma_1^c = 0.387$ GeV.

\begin{figure}[!t]
\begin{center}
\parbox{15cm}{
\psfig{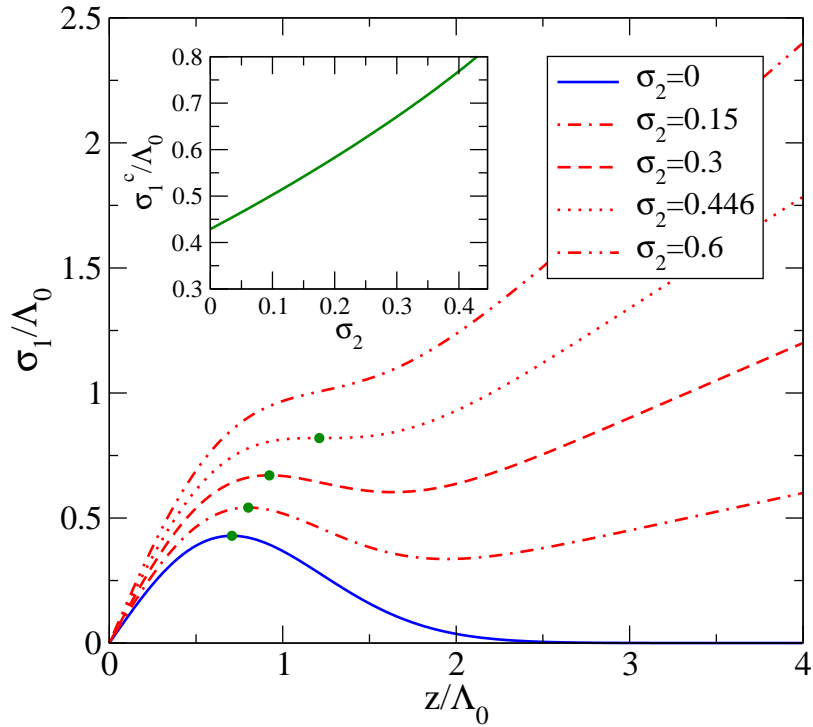}
}
\caption{(Color online) The structure of only real
singularities $z$ as functions of the mass gap $\sigma_1$ are 
shown for the Gaussian regulators of a rank-2 model.
Setting $\sigma_2=0$ leads back to the rank-1 model given by the
blue curve.
Non-zero values of $\sigma_2$ then give a family of red curves, where
the green dot gives $\sigma_1^c(\sigma_2)$ (see text).
The full function $\sigma_1^c(\sigma_2)$ is obtained numerically
and shown in the inset.
}
\label{rank2_crit}
\end{center}
\end{figure}
For rank-2 we facilitate the analysis further by considering the case
$\Lambda_0 = \Lambda_1$.
With $\sigma_2=0$ two real poles exist, as 
shown on Fig.~\ref{rank2_crit}.
Any $\sigma_2>0$ brings an extra pole $\sigma_1/\sigma_2$ 
from infinity.
As $\sigma_2$ increases, this singularity in turn coalesces with
the first two at $\sigma_2=\sigma_2^c = 2/e^{3/2}\simeq 0.446$,
after which point only one real singularity is present for all
values of $\sigma_1$.
At the same time, the threshold $\sigma_1^c$ rises as a 
function of $\sigma_2$, until it reaches
\begin{equation}
\sigma_1^c(\sigma_2^c) = 
\frac{\Lambda_0}{\sqrt{2}}\left(\frac{3}{e}\right)^{3/2}~,
\end{equation}
as shown by the green line
on the inset of Fig.~\ref{rank2_crit}. 
The outcome is that in rank-2 it is 
easier for the physical
mass gap to be undercritical.
A concrete calculation for set B with $\Lambda_0=\Lambda_1$ 
yields $\sigma_1=0.497$ GeV, $\sigma_2 = 0.430$, so that $\sigma_1^c(0.430)=0.652$ GeV 
confirming that indeed
the gap is undercritical.

\subsection{Entropy density}

\begin{figure}[!t]
\begin{center}
\parbox{15cm}{
\psfig{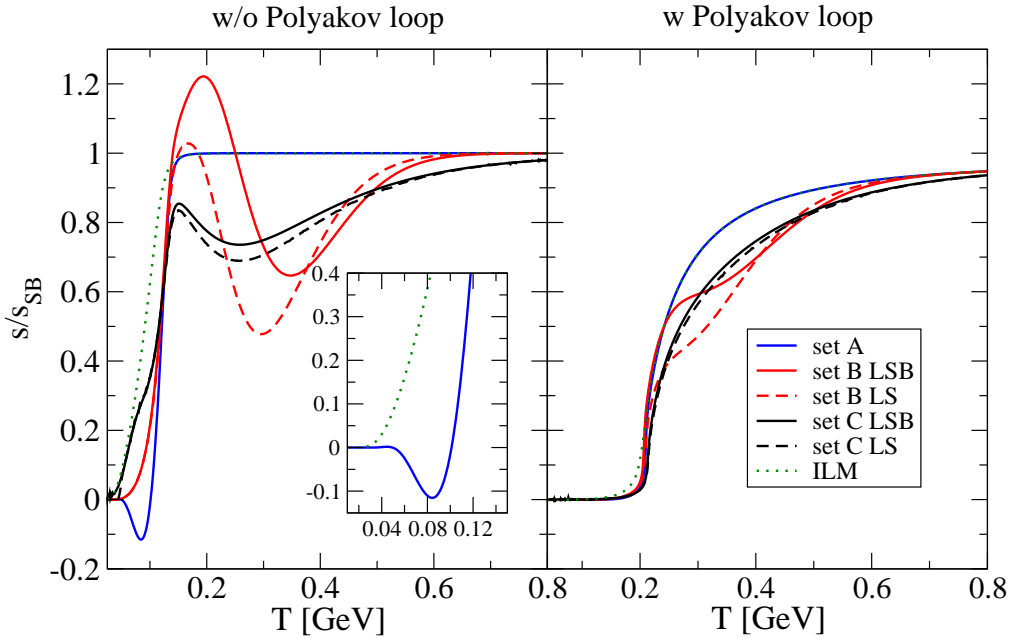}
}
\caption{(Color online) Scaled entropy density as a function of temperature.
Left panel is without PL, while the right panel includes the PL.
Notice the temperature mismatch in the oscillations for set A versus 
sets B and C.
Since the gap in the ILM model is undercritical, there are no oscillations
in the low $T$ phase even in the case without the PL, as shown in the inset.}
\label{entr}
\end{center}
\end{figure}
At this point we are ready to analyze the resulting mean 
field EoS at finite temperature as obtained
from
\be
\label{pressure}
p = -\Omega_\mathrm{reg}~.
\ee
It is particularly useful to examine the entropy density
\be
s = \frac{dp}{dT}~.
\ee
Being a derivative of the pressure (\ref{pressure}) w.r.t.
the temperature, the entropy density will make
any possible unphysical behavior most transparent,
such as the oscillations found in \cite{Benic:2012ec}, and
therefore be suitable for selecting a preferable model.

If the gap is overcritical, oscillations shall 
be present in Gaussian models;
the complex exponential of the regulators
giving rise to an infinite number of poles.
It is more involved to extract analytic structure of set C,
containing cuts as well as poles, so
we restrict our discussion to the numerical results.
In all cases, the PL will play an important role.
The results in Fig.~\ref{entr} are given for
all three different regulators and for the ILM model 
and scaled to the massless Stefan-Boltzmann
(SB) value.
Whereas a smooth, monotonous rise in the entropy is expected
as the quark degrees of freedom are 
liberated, non-physical oscillations
are present for all three regulators, as anticipated in the
first subsection.

In order to underline the fact that complex singularities are crucial
for oscillations it is useful to 
consider the comparison the entropy in set A (blue)
and in the ILM (green, dotted curve).
The low $T$ region is shown in the inset on the left panel.
There it is clearly visible that the effect of the CCMPs
is given in the low temperature region for set A, but not for the ILM.
The reason is that since the 
mass gap in the latter case is undercritical, the lowest lying singularities,
which dominate the entropy at low $T$, are real, see also Fig~\ref{rank2_crit}.

Furthermore, the SB limit is well saturated already at $T\gtrsim T_c$
when the system is not coupled to the PL, see the blue curve on the
left panel of Fig.~\ref{entr}.
A significant change in the onset is achieved when 
coupling to the PL,
but this should be attributed to the fact that the PL potential
$\mathcal{U}(\Phi)$
is fitted to lattice data for the pressure of pure glue.

For set B, which in addition has the WFR channel, the mass gap
becomes undercritical (see Fig.~\ref{rank2_crit}),
so that the behavior of entropy is 
monotonous at $T\lesssim T_c$ as observed by the
red curve on the left panel of Fig.~\ref{entr}.
In contrast, here the oscillatory behavior 
is present exclusively at $T\gtrsim T_c$.
Due to the analysis in the first subsection we may again attribute
this behavior to complex singularities.
But, since $\sigma_B$ is 
drastically reduced,
they are linked
to the analytical properties of the WFR term.

In order to confirm this conjecture, it is sufficient to look for
complex poles for set B in a idealized scenario where the
mass gap is zero, and where $A(p^2)=C(p^2)$.
If we are able to prove that there are poles in 
the degenerate quark propagator (\ref{eq:qprls}) 
with  $B(p^2)=0$, besides
the massless one, then
we can use Eq.~(\ref{eq:osc}) to again argue that 
they are responsible
for oscillations seen in Fig.~\ref{entr}.
For set B, one can show that the 
condition $A(-\mathcal{E}_k^2)=0$, where $\mathcal{E}_k = \epsilon_k + i\gamma_k$,
is fulfilled with
\be
\epsilon_k(\mathbf{p}) = 
\frac{\Lambda_1}{\sqrt{2}}\left[\sqrt{\left(\frac{\mathbf{p}^2}{\Lambda_1^2}-\log \sigma_2\right)^2+(2k+1)^2\pi^2}+\frac{\mathbf{p}^2}{\Lambda_1^2}-
\log \sigma_2\right]^{1/2} \, ,
\qquad \gamma_k = \frac{(2k+1)\pi}{2\epsilon_k}\Lambda_1^2~,
\label{eq:pol_sB}
\ee
where $k\in \mathbb{Z}$ and for set C
\be
\mathcal{E}_k(\mathbf{p}) =
\Lambda_1\left[1+\frac{\mathbf{p}^2}{\Lambda_1^2}-
(\alpha_z+\sigma_2+\alpha_z \sigma_2)^{2/5}
e^{\frac{2\pi k}{5}i}\right]^{1/2}~,
\label{eq:pol_sC}
\ee
with $k=0,1,...,4$.
Interestingly, in set B, even though the number of poles is infinite,
we can still find a clear hierarchy.
For example, if $\sigma_2=1$, then 
$$m_k^R=m_k^I=\Lambda_1\sqrt{\frac{\pi}{2} + k\pi}~.$$
Notice also that as $\sigma_2\to 0$, for set B we 
find $\epsilon_k(\mathbf{p})\to\infty$ and $\gamma_k(\mathbf{p})\to 0$, ensuring
that in the high temperature range where $\sigma_2\to 0$ one is left
with the usual massless singularity.
The analogous formula for set C (\ref{eq:pol_sC}) is valid only when
$\sigma_2\neq 0$: the limiting case is provided by going back to the original
formula $p^2 A^2(p^2)=0$.
More important, as (\ref{eq:pol_sB}) and (\ref{eq:pol_sC}) 
are double poles, the SB limit is eventually exceeded,
as demonstrated by the red curve in Fig.~\ref{entr}.
This unsatisfactory result is readily 
improved with the lattice adjusted set
C parametrization; the oscillation
is somewhat reduced, giving an entropy 
within the SB bound, over the whole
temperature range.

Introducing the PL to the system leads to a 
dramatically improved behavior.
As the right panel of Fig.~\ref{entr} indicates, there is a
smooth rise in the entropy for set A, in accordance
with (\ref{pkinpl2}).
The PL is very successful in taming the oscillations
in a theory with CCMPs, as its value is zero in the low temperature,
confined phase.
As the confinement transition is coincident with the chiral one,
the only poles that the PL 
is able to strongly suppress are the ones present
before the chiral transition.
Therefore, the oscillation in 
set B, due to the double poles, is still present,
albeit largely reduced, owing to the fact that $\Phi$
is still less than unity in that region.
For set C, the oscillation was smaller to begin with,
so when the PL smooths
that out, all what is left is again 
a monotonous rise, as observed by the black
curve on the right panel of Fig.~\ref{entr}.
The same effect is visible in a recent
calculation in $N_f=2+1$ nl-PNJL \cite{Carlomagno:2013ona}.

\subsection{ Influence of Lorentz symmetry breaking}
The influence of LSB is minor, being somewhat stronger for set B.
In particular, the two black curves on the right panel of Fig. \ref{entr}
for entropy density in set C with PL are almost identical, whereas for set B, LSB
can lead even to a 20\% increase for $T\gtrsim T_c$.
A qualitative understanding of this effect can be achieved from the
quasi-particle picture given by (\ref{disp}).
The particular value of the entropy could be seen as the interplay of the two
effects: increasing $m_\mathrm{qp}$ decreases the entropy (``loss" term),
while increasing
$v_\mathrm{qp}$ increases the entropy (``gain" term).
A ratio of the masses and the velocities for the LSB and the LS case, yields
\be
\frac{v_\mathrm{qp}^\mathrm{LSB}}{v_\mathrm{qp}^\mathrm{LS}} =
\frac{1+\sigma_A}{1+\sigma_C} \, , \quad
\frac{m_\mathrm{qp}^\mathrm{LSB}}{m_\mathrm{qp}^\mathrm{LS}} =
\frac{1+\sigma_2}{1+\sigma_C} ~,
\ee
where we have used that $v_\mathrm{qp}^\mathrm{LS}=1$, and
$\sigma_1 \simeq \sigma_B$, which is
well fulfilled in our case, see Fig. \ref{lorentzac}.
From Fig. \ref{lorentzac} we also deduce
that $\sigma_2>\sigma_A$, thus
\be
\frac{m_\mathrm{qp}^\mathrm{LSB}}{m_\mathrm{qp}^\mathrm{LS}} :
\frac{v_\mathrm{qp}^\mathrm{LSB}}{v_\mathrm{qp}^\mathrm{LS}} =
\frac{1+\sigma_2}{1+\sigma_A} > 1~,
\ee
which can be interpreted to mean that the loss term in the entropy density
is less significantly affected by LSB than the gain term, providing a
net increase of the entropy density.

\section{Meson decay widths at finite temperature}
\label{sec:mes}
At this point we discuss the thermal behaviour of 
mesonic degrees of freedom.
In a local NJL setup this has been thoroughly studied.
We expect that non-local interactions might induce new features particularly
into the picture of meson dissociation in the plasma.
The aim is to deduce qualitative influence of 
non-local interactions 
on the aspect of Mott physics such as 
resonance broadening, and also to discuss the effects of
the WFR channel.
Since the explicit
calculations are performed with LSB all the mean fields are
denoted as $\sigma_{A,B,C}$.

The in-medium features of correlations are encoded in the meson
polarization function
\cite{Blaschke:2000gd,Scarpettini:2003fj,Horvatic:2007wu,Contrera:2009hk}
\be
\Pi_M (\nu_m,|\mathbf{q}|)=
\frac{8N_c}{3} 
T\sum_{n=-\infty}^\infty \int\frac{d^3 p}{(2\pi)^3}
\mathrm{tr}_C\left[g^2(\tilde{p}_n^2)
\frac{K_M(\tilde{\omega}_n^2,\mathbf{p}^2,\nu_m^2,\mathbf{q}^2)}
{\mathcal{D}((\tilde{\omega}_n^+)^2,(\mathbf{p}^+)^2)
\mathcal{D}((\tilde{\omega}_n^-)^2,(\mathbf{p}^-)^2)}\right]~,
\label{pol}
\ee
with
\be
\begin{split}
K_M(\tilde{\omega}_n^2,\mathbf{p}^2,\nu_m^2,\mathbf{q}^2)=
(\tilde{\omega}_n^+ \tilde{\omega}_n^-)&
C((\tilde{p}_n^+)^2)C((\tilde{p}_n^-)^2)\\
&+
(\mathbf{p}^{+}\cdot\mathbf{p}^{-})A((\tilde{p}_n^+)^2)A((\tilde{p}_n^-)^2)\pm
B((\tilde{p}_n^+)^2)B((\tilde{p}_n^-)^2)~,
\end{split}
\label{gem}
\ee
generalized in order to include effects of LSB.
We use the subscript $M$ for specifying the meson $M=\pi,\sigma$ and denote
the meson 4-momentum as $q_m =(\nu_m,\mathbf{q})$, where
$\nu_m = 2m\pi T$ are the bosonic Matsubara frequencies.
Furthermore, $\tilde{p}_n^{\pm} = (\tilde{\omega}_n^{\pm},\mathbf{p}^\pm)$,
with $\tilde{\omega}_n^{\pm} = \tilde{\omega}_n \pm \nu_m/2$, and
$\mathbf{p}^\pm = \mathbf{p}\pm \mathbf{q}/2$ and
\be
\mathcal{D}(-z^2,\mathbf{p}^2)=
\mathbf{p}^2A^2(-z^2+\mathbf{p}^2)-z^2 C^2(-z^2+\mathbf{p}^2)+
B^2(-z^2+\mathbf{p}^2)~.
\label{denom}
\ee
It will be crucial to note that Eq.~(\ref{pol}) is 
valid only for sets A -- C.
The polarization function in the ILM follows by making a replacement
\be
g^2(\tilde{p}_n^2) \to r^2((\tilde{p}_n^+)^2)r^2((\tilde{p}_n^-)^2)~,
\label{eq:repl}
\ee
in the first term after the square bracket in Eq.~(\ref{pol}).
The regulator in the $B(p^2)$ function in the propagator 
is altered accordingly, i. e. so that $g(p^2) \to r^2(p^2)$.

\subsection{Meson widths}

The width is obtained by renormalizing the meson propagator.
For simplicity, if we take the vacuum propagator in Euclidean space
and expand it around $q^2 = -m_M^2$
\be
\Delta(q^2) = \frac{1}{-\frac{1}{G_S}+\Pi_M(q^2)} \to
\frac{g_{M\bar{q}q}^2}{q^2+m_M^2+i\Gamma_M m_M}~,
\ee
where $\Gamma_M$ is the meson width
\be
\Gamma_M = g_{M\bar{q}q}^2\frac{\mathrm{Im}(\Pi_M)}{m_M} ~,
\label{wid}
\ee 
and $g_{M\bar{q}q}$ is the effective quark-meson coupling, or the meson wave
function renormalization
\be
g_{M\bar{q}q}^2 =
\left[\frac{\partial\mathrm{Re}(\Pi_M)}
{\partial q^2}\right]^{-1}_{q^2 = -m_M^2}~.
\label{eq:qqm}
\ee

We will obtain $\mathrm{Im}(\Pi_M)$ as a function of 
the meson energy, denoted by $q_0$, at rest $\mathbf{q}=0$.
In doing so, we will use several simplifications and approximations,
to be stated precisely in the following.

First of all, it is known in the 
literature \cite{Cutkosky:1969fq,Bhagwat:2002tx,Scarpettini:2003fj}
that an elaborate analytic continuation of the 
polarization loop
is possible which does not lead to thresholds in the case
the singularities of the quark propagator are complex.
In other words, if the quark propagator has only complex
singularities, the meson is stable.

As we have shown, some of the models that we study here, like
set A, have such property in the vacuum.
On the other hand, models like set B and ILM
have also real singularities in the vacuum.
Taking the parameters given in Table \ref{tab:parameters}, their values,
denoted as $m_L$, 
are $m_L=0.508$ GeV for set B and $m_L=0.331$ GeV for ILM. 
So, in principle, if the condition for the kinematic threshold
is satisfied, i. e. if $m_M>2m_L$ the meson must be unstable.
It turns out that for set B and ILM this is not the case - in other
words, pion and sigma mesons are stable in the vacuum, the explicit
values are collected in Table \ref{mott_tab}.

\begin{figure}[!t]
\begin{center}
\psfig{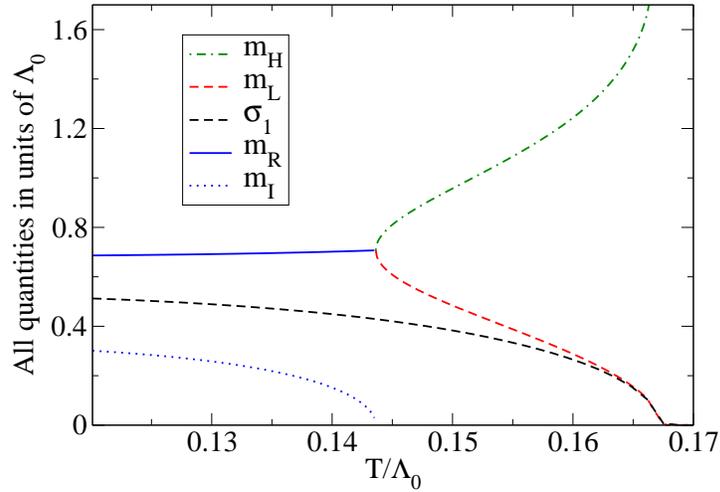}
\caption{(Color online)
Temperature dependence of lowest singularities for set A in the chiral limit
and without Polyakov loop.
The full blue, and dotted blue line are the real ($m_R$) and 
imaginary ($m_I$) part of
the lowest lying singularity.
Beyond a certain temperature given by the condition $\sigma_B(T) = \sigma_B^c$,
these poles join on the real axis to form two real poles, $m_L$ given by
the red dashed line, and $m_H$ given by the green dash-dotted line.
The mass gap $\sigma_B$ is also shown to illustrate how as $T\to T_c$, 
$m_L$ approaches $\sigma_B$.}
\label{sings_fig}
\end{center}
\end{figure}

Proceeding to finite $T$ it is possible for mesons to develop finite
imaginary parts if some of kinematic thresholds become allowed.
A complete discussion requires mapping the behavior of the singularities 
as a function of temperature, which in turn
requires mapping 
them as a function of the mean fields $\sigma_{A,B,C}$.
In the case of rank-1 models, like set A and ILM, we have a single mean field $\sigma_B$.
Then the thermal dependence of the lowest
lying singularities can be numerically mapped, and are shown 
on Fig.~\ref{sings_fig} for set A in the chiral 
limit, where singularities are complex in
vacuum.
As the temperature increases, $\sigma_B$ decreases - when it reduces
below $\sigma_B^c$ we have real singularities, denoted as $m_L$ and $m_H$.
The singularity $m_H$ becomes very heavy as we approach chiral
restoration - therefore the meson will not decay into the state $m_H$.
On the contrary, $m_L$ becomes the massless, chiral singularity,
therefore we consider the decay of the meson to $m_L$.

For set B and set C models, where
additional mean fields are present, although this is in 
principle possible, we do not map the singularities
as functions of the mean fields.
Based on the previous analysis in the vacuum, and for
rank-1 models also at finite temperature, we 
anticipate the following idealized scenario.
At low temperatures all the singularities in the
models are either complex or real, but in 
both cases they are at least of the
order of the scale of the 
regulators $f(p^2)$ 
and $g(p^2)$, which is
out of reach as a continuum threshold.
Increasing the temperature, the mass gap drops.
This forces one pole to proceed to $m_\mathrm{qp}$ as defined
in Eq.~(\ref{disp}) and then to the origin in the complex
plane becoming the physical, current quark mass for very
high temperatures.
See
Fig.~\ref{sings_fig} for an explicit
example in the case of rank-1.
The other auxiliary states have either complex masses, or very heavy
real masses.
In either case it is important to realize that they will not
contribute to the imaginary part.
In total, the imaginary part, and therefore, non-zero width, will be generated by the decay of the meson to the singularity that
continuously evolves to the current quark mass.

Now we can calculate the imaginary part by applying
the $i\epsilon$ prescription for the mass $m_\mathrm{qp}$

\be
\mathrm{Im}[\Pi_M(-iq_0,0)] = \frac{1}{2i}
\left[\Pi_M(-i(q_0+i\epsilon),0)-\Pi_M(-i(q_0-i\epsilon),0)\right]~,
\label{immeson}
\ee
where the bosonic Matsubara frequencies were analytically continued to
$i\nu_m \to q_0$, and where the imaginary part will be calculated at $\mathbf{q}=0$.
The master formula for performing the summation over the fermionic Matsubara
frequencies, as well as the
detailed derivation of the imaginary part of
(\ref{pol}) are collected in the Appendix \ref{app}.
Here we quote the final result for sets A -- C
\be
\begin{split}
\mathrm{Im}[\Pi_M(-iq_0,0)] = \frac{d_q}{16\pi}
&\left[1- n_+^\Phi(q_0/2)-n_-^\Phi(q_0/2)\right]
\sqrt{1-\left(\frac{2m_\mathrm{qp}}{q_0}\right)^2}\\
&\times g^2\left(\frac{q_0^2}{4}-m_\mathrm{qp}^2\right)
\frac{K_M\left(0,\frac{q_0^2}{4}-m_\mathrm{qp}^2,-q_0^2,0\right)}
{\left[\mathcal{D}'
\left(-\frac{q_0^2}{4},\frac{q_0^2}{4}-m_\mathrm{qp}^2\right)\right]^2}
\theta\left(\frac{q_0}{2}-m_\mathrm{qp}\right)~,
\end{split}
\label{img_pol}
\ee
with $\mathcal{D}'$ defined by (\ref{dden}).
The square bracket in the first line of (\ref{img_pol}) defines the
Pauli blocking term, with $n_\pm^\Phi(z)$ being the generalized occupation
number for fermions in the presence of the Polyakov loop $\Phi$ and its
conjugate $\bar{\Phi}$,
\be
n^\Phi_\pm(z) = \frac{\bar{\Phi}e^{-\beta(z\mp\mu)}+
2\Phi e^{-2\beta (z\mp\mu)}+e^{-3\beta(z\mp\mu)}}
{1+3\bar{\Phi}e^{-\beta(z\mp\mu)}+3\Phi e^{-2\beta(z\mp\mu)}
+e^{-3\beta(z\mp\mu)}}~.
\label{occ}
\ee

The imaginary part of the polarization loop
$\mathrm{Im}(\Pi_M)$ for ILM follows by making the 
replacement (\ref{eq:repl}) while taking into account that the quasi-particle 
energies are dictated by energy conservation, see
the $\delta$-function in Eq.~(\ref{discc}), yielding
\be
g^2\left(\frac{q_0^2}{4}-m_\mathrm{qp}^2\right)\to r^4(-m_\mathrm{qp}^2)~.
\label{eq:rep_pol}
\ee
Notice that as the quasi-particle mass goes to the current quark mass, in the ILM
model this prefactor $r^4(-m_\mathrm{qp}^2)\to 1$.
On the other hand, in sets A -- C, ignoring the small current mass, 
we will still be left with $g^2\left(q_0^2/4\right)$. 
This might have a significant impact in the high $T$ 
phase, depending on the value of $q_0$.

It is interesting to discuss the local limit, where we obtain
\be
K_M\simeq \frac{q_0^2}{4}C_0^2 +
\left(\frac{q_0^2}{4}-m_\mathrm{qp}^2\right)A_0^2\pm m_\mathrm{qp}^2
\, , \qquad \mathcal{D}'\simeq C_0^2~.
\label{gm_njl}
\ee
Furthemore, by taking $A_0,C_0\to 1$, we reproduce
the local NJL result \cite{Hansen:2006ee}
\be
\begin{split}
\mathrm{Im}[\Pi_M(-iq_0,0)] \to \frac{d_q}{16\pi}&
\left[1-n_+^\Phi(q_0/2)-n_-^\Phi(q_0/2)\right]\\
&\times\sqrt{1-\left(\frac{2m_\mathrm{qp}}{q_0}\right)^2}
\left[\frac{q_0^2}{4}-m_\mathrm{qp}^2\pm m_\mathrm{qp}^2\right]
\theta\left(\frac{q_0}{2}-m_\mathrm{qp}\right)~.
\end{split}
\label{img_pol_njl}
\ee
On the other hand, by using (\ref{gm_njl}), in the chiral limit we obtain
\be
\frac{K_M}{(\mathcal{D}')^2}\to \frac{q_0^2}{4}\frac{1}{C_0^2}
\left(1+v_\mathrm{qp}^2\right)~.
\label{lsb_pol}
\ee
This result shows that introducing WFR
can significantly reduce the imaginary part.
In addition, if LSB by the medium is acknowledged,
owing to the fact that $v_\mathrm{qp}<1$, the imaginary 
will be even more reduced.

As we expect degeneracy of meson states above the chiral
transition temperature, in practice it will be sufficient to
consider the pion width.
In order to do that we need two more ingredients:
$q_0$ and $g_{\pi\bar{q}q}$.
In local NJL, see e. g. \cite{Hansen:2006ee} and 3D nl-NJL 
studies \cite{Schmidt:1994di}, it is 
shown that $g_{\pi\bar{q}q}$ is a slowly 
varying function of the temperature.
Actually, for $g_{\pi\bar{q}q}$ this can be naturally understood
from the quark-level Goldberger-Treiman relation
$g_{\pi \bar{q}q}\sim m_\mathrm{qp}/f_\pi$ where $f_\pi$ is the pion decay constant.
Up to the temperatures close to chiral restoration both $m_\mathrm{qp}$ 
and $f_\pi$ are constant,
while around and after $T_c$ they both get monotonously reduced. 
Hence, to get an idea on 
the pion width also in a covariant (4D) nl-PNJL setup, 
we make a rough approximation by replacing the thermal
dependence of the 
quark-pion coupling $g_{\pi\bar{q}q}$ by its vacuum value.
The calculated values for all the models considered in this work
are collected in Table \ref{mott_tab}.

\begin{table}[t]
\begin{tabular}{| c | c | c | c | c |}
\hline
-- & set A  & set B & set C & ILM \\ \hline
$\sigma_B$ [GeV] & 0.424 & 0.429 & 0.442 & 0.284\\ \hline
$\sigma_B^c$ [GeV] & 0.317 & 0.557 & 0.0 & 0.391 \\ \hline
$m_L$ [GeV] & --  & 0.508 & -- & 0.330 \\ \hline
$T_\mathrm{cont}$ [GeV] & 0.208 & 0 & 0 & 0 \\ \hline
$T_\mathrm{Mott}^\pi$ [GeV] &  0.21 & 0.21 & 0.22 & 0.20 \\ \hline
$T_\mathrm{Mott}^\sigma$ [GeV] & 0.21 & 0.21 & 0.21 & 0.20\\ \hline
$m_\pi$ [GeV] & 0.14 & 0.14 & 0.14 & 0.14 \\ \hline
$m_\sigma$ [GeV] & 0.68 & 0.63 & 0.56 & 0.4 \\ \hline
$g_{\pi\bar{q}q}$ & 4.62 & 5.74 & 4.74 & 2.47 \\ \hline

\end{tabular}
\caption{For sets A--C, and the ILM, the table collects
vacuum values of the mass gaps $\sigma_B$, the
critical values $\sigma_B^c$ at which the physical continuum moves from the
real axes, together with the respective temperature $T_\mathrm{cont}$
where this happens.
We also provide the Mott temperatures for $\pi$ and $\sigma$ mesons.
Note that for set A the physical mass gap is overcritical, while for set B and the ILM it
is undercritical.
For set C, the imaginary part develops continuously from the current quark
mass $m$.
Therefore, the continuum is present in ILM, and sets B 
and C already at $T=0$.
The table also gives the actual values for the lowest 
real singularities
in the vacuum, denoted by $m_L$.
Furthermore, we provide the vacuum values of the masses and the
quark-pion couplings.}
\label{mott_tab}
\end{table}

\subsection{Meson masses}

To calculate the width we still need $q_0$ which should in principle
be given by the dynamical, pole mass $m_M^\mathrm{pole}$, obtained from its Bethe-Salpeter equation
at zero meson momentum $\mathbf{q}=0$
\be
1-G_S\Pi_M(-im_M^\mathrm{pole},0) = 0~.
\ee
While such calculations are straightforward in local NJL models, the covariant
approach presents technical difficulties.
Namely, a complete analysis requires performing Matsubara summation analytically. 
Since the polarization loop contains a pair of quark
propagators, via residue calculus, this will in principle lead to
a double summation over all the singularities present in the
propagator, requiring that their behavior first needs to be traced as a function
of the mean fields $\sigma_{A,B,C}$.
Note that this is significantly more 
involved than the imaginary part
since here we need the information on singularities in the low as well
as in the high $T$ regime, whereas for the imaginary
part we needed only one singularity in the high $T$ regime.

Since the aim of the present section 
is the qualitative analysis of the meson widths obtained 
within models,
for $q_0$ we have chosen to use by hand the 
meson screening masses \cite{Blaschke:2000gd,Scarpettini:2003fj,
Horvatic:2007wu,Contrera:2009hk} $m_M^\mathrm{spat}$ given by solving the equation
\be
1-G_S\Pi_M(0,-im_M^\mathrm{spat}) = 0~.
\label{eq:spat_eq}
\ee 
This simplification is supported by a calculation 
in local NJL models \cite{Florkowski:1993bq,Florkowski:1993br} 
where a careful comparison 
of both screening and pole masses lead to the following 
conclusion: at low temperatures, below the chiral 
restoration temperature, the screening masses 
closely follow the dynamical ones. 
However, at temperatures above the chiral 
restoration, screening masses were found to be 
somewhat higher in value. 
It should be emphasized that both the screening 
and the pole masses were found to 
follow the expected pattern of chiral 
symmetry breaking and restoration.
\begin{figure}[t]
\begin{center}
\psfig{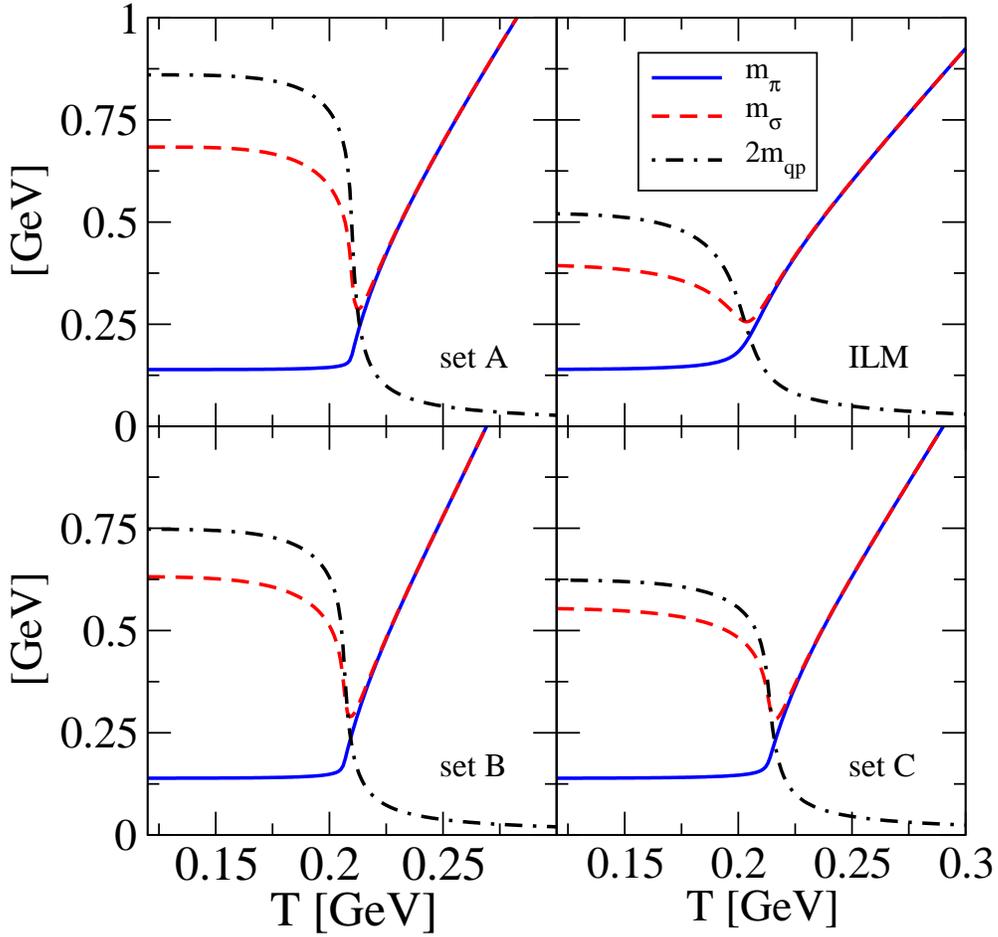}
\caption{(Color online)
The panels display screening masses  
for $\pi$ and $\sigma$ mesons
for different sets.
The results for sets B and C are only for LSB case.}
\label{mesons_fig}
\end{center}
\end{figure}

\subsection{Discussion of the results}

\begin{figure}[t]
\begin{center}
\psfig{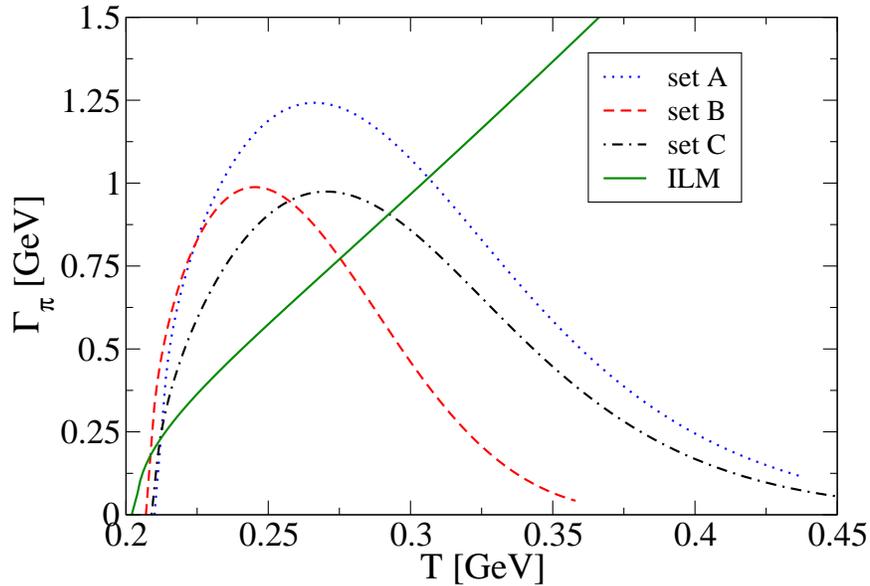}
\caption{(Color online)
The figure displays the approximate pion widths 
calculated from Eqs.~(\ref{wid}) 
for sets A -- C, and ILM.}
\label{widths_fig}
\end{center}
\end{figure}
On Fig.~\ref{mesons_fig} we 
plot the sigma and pion spatial masses,
as calculated from Eq.~\ref{eq:spat_eq}.
Besides the spatial meson masses, it is instructive to 
show the ``continuum'' states defined by $2m_\mathrm{qp}$, where
$m_\mathrm{qp}$ is given by (\ref{disp}).
Strictly speaking, these states need not be present 
as actual singularities of the
quark propagator up to some high temperature, as 
was previously discussed.

Returning to our canonical example in set A, the 
continuum states are developed
only after the temperature where $\sigma_B = \sigma_B^c$.
For finite current quark mass, this happens at $T_\mathrm{cont} = 0.208$ GeV.
Let us now define the Mott temperatures by
$$m_M^\mathrm{spat}(T_M^\mathrm{Mott}) = 2m_\mathrm{qp}(T_M^\mathrm{Mott})~.$$
Now, from Fig.~\ref{mesons_fig} we observe that the Mott temperatures for both
$\pi$ and $\sigma$ are higher than $T_\mathrm{cont}$, {\it i.e.},
$T^\mathrm{Mott}_\pi = 0.213$ GeV, $T^\mathrm{Mott}_\sigma = 0.212$ GeV,
thus providing a picture where the continuum of states
should be first realized in the singularities of the quark
propagator, so that the meson decay can happen only at higher
temperatures.
This is also the situation in all 
other models, i. e. $T^\mathrm{Mott}_M > T_\mathrm{cont}$ for 
sets B and C
and ILM.
The complete set of values of Mott and continuum temperatures is
collected in Table~\ref{mott_tab}.

We see that introducing WFR lowers the continuum
according to the Eq.~(\ref{disp}).
Also the $\sigma$ meson mass is reduced, which one would naively agree to
from the PNJL setting where $m_\sigma \simeq 2 m_\mathrm{qp}$.
The meson screening masses are joining at the chiral restoration temperature,
and tend to rise steeply beyond that point, approaching $1$ GeV already
around $T\simeq 0.3$ GeV, with the steepest rise for set B.
The results for the ILM model single out because of its small
mass gap, which in turn leads to a smoother transition into 
the chirally restored phase.
As a further consequence, the sigma meson mass is almost twice
reduced in the vacuum.

We calculate the widths by using spatial masses in Eq.~(\ref{wid}), 
and in Eq.~(\ref{img_pol}) by replacing $q_0 \to m_M^\mathrm{spat}$, instead
of the more accepted $m_M^\mathrm{pole}$.
This certainly introduces an error in our calculation, but
since the qualitative behavior
of both spatial and dynamical masses is the same it 
will nevertheless provide a valuable study.
In that sense our results will be best seen as a study of the thermal
dependence of the imaginary part of the polarization loop.
Namely, instead of using a phenomenological fitting function for the mass as a function
of the temperature, we employ the calculated screening masses.
Then, since the width and the imaginary part are proportional, 
the \textit{convention} for calculating the width itself is motivated by
the fact that we would like to interpret our results physically.

Fig.~\ref{widths_fig} shows the main result of this Section.
For sets A -- C, inspired by the separable DSE calculation, the widths
follow a generic pattern.
In the low temperature region, we find a steep rise, mostly due to the meson mass
itself: see e. g. the local and the chiral limit (\ref{lsb_pol}), where 
one has a quadratic
dependence on the meson mass in the imaginary part of the polarization loop,
giving a linear slope for the width, see Eq.~\ref{wid}.
But, since in the non-local models, the complete imaginary part, and therefore
the width is multiplied by the regulator, it is this factor that dictates
the high temperature behaviour.
Namely, as $g(p^2)$ is a rapidly decreasing function of 
momenta (see Eqs.~(\ref{eq:setA})-(\ref{eq:setC})),
and because for sets A -- C, the argument is a rising function of the temperature,
it eventually overwhelms the quadratic dependence, and provides a characteristic
decrease in the width.
Therefore, in the high temperature phase, the width drops to zero.
The quantitative result shown on Fig.~\ref{widths_fig} might be somewhat 
exaggerated due to the fact that the screening
masses steeply rise with the temperature, making the decline of $\Gamma_{\pi}$
more dramatic.
Nevertheless, the qualitative 
behavior should be considered as generic to this 
class of models.
In that regard, let us also comment on the fact that, as announced in the
previous subsection, the width in overall gets somewhat reduced
when the WFR channel is introduced.
This is demonstrated by the dashed, red and dash-dotted, black curves in Fig.~\ref{widths_fig}.

Concentrating on the ILM calculation 
of the width, the result we obtain is completely
different: due to the fact that the regulator in this case has a different momentum
dependence in the polarization 
loop, see Eq.~(\ref{eq:repl}) and Eq.~(\ref{eq:rep_pol}), there is no
dependence on the meson mass in the regulator, and its effect at high temperatures
is highly suppressed.
This results in a monotonous rise of $\Gamma_\pi$, shown 
by the full, green curve, in
the low, as well as in the high temperature region.

\section{Conclusions}
\label{sec:concl}

In this work we have discussed a class of nonlocal 
PNJL models which are
suitably adjusted to model the behaviour of the quark propagator in the vacuum
as determined in lattice QCD simulations.
These are extrapolated to finite $T$ and $\mu$ whereby 
the new element of medium induced
Lorentz symmetry breaking is introduced.
In Sec.~\ref{sec:vac_crit} we have examined 
the influence of this term on the phase diagram,
in the mean field approximation.
While LSB provides a significant difference in the wave function
renormalization channel mean fields
after $T_c$, we conclude
the critical properties and the EoS do not change appreciably.
We find in general that models with WFR tend to slightly lower the
position of the CEP on the critical line.
Complementary to numerical results, a thorough analytic study
of the critical behavior in the vacuum and in the medium was given.
Where possible, analytical limits to the local PNJL model were also given.
While these are only estimates, it might be interesting to also examine a
non-trivial WFR in a complete numerical setup of local PNJL.

In Sec.~\ref{sec:eos} we have calculated 
the EoS concluding that for a wide 
class of nl-NJL models the EoS is oscillatory.
We have demonstrated that in contrast to rank-1 models with
Gaussian regulator, for rank-2 models with Gaussian regulators, the mass
gap is undercritical, thus giving a mismatch in the temperature where the
oscillations in the EoS occur.
While for rank-1 they occur in the chirally broken phase, in rank-2 they occur
in the chirally restored phase.
For Lorentzian regulators, as in set C, we have found that the oscillations are
also present, but somewhat less drastic.
Such oscillations violate general thermodynamic criteria
for stability of the system, and are not observed in lattice calculations. 
We have found that an improvement of the gluon sector, e. g. in the form
of the Polyakov loop, significantly improves the thermodynamics.
Nevertheless, since the Polyakov loop is finite in the high temperature
phase, the oscillations in rank-2 models are only reduced.

In Sec.~\ref{sec:mes} we have 
presented a
detailed derivation and a
discussion of the widths in the 
covariant version of nonlocal models.
We emphasize that the latter
was completely absent from the literature, although
the model itself is present in the community 
for more than two decades.
The basic problem is the covariance of the approach.
More precisely, the fact that it is defined in Euclidean space,
makes ``Minkowski-quantities'' like the dynamical
meson masses and widths, difficult to
obtain.
Since we do not claim that we have 
solved this hard problem\footnote{A first step in solving it would be
to map the analytic structure of the quark propagator in the complex plane.
This is a highly non-trivial task, addressed only 
very recently \cite{Strauss:2012dg,Windisch:2012zd,Windisch:2012sz}.}, 
the main drawback being that we have not calculated the dynamical
pole masses, but the spatial ones,
we are nevertheless of the opinion that the results that we do display
are still interesting to the community, as they 
bear a qualitative significance.

Thus, given the roughness of our approximations we can state the following.
First, the meson widths, as calculated in our approximation are not
strongly affected by the shape of the regulator that is used.
Second, introducing WFR and LSB reduces the widths to some extent,
and third; the most interesting result comes from investigating the different
ways non-local interactions can be introduced.
For sets A -- C, where the non-locality is inspired by separable
DSE model, the widths rapidly decline at high temperatures.
On the other hand, if the non-locality is introduced via ILM
the width is a rising function of temperature.
It should be noted that the latter result is also similar to what is seen in
local \cite{Hansen:2006ee}, or 3D non-local \cite{Schmidt:1994di} NJL studies.

Future studies should
acknowledge that, after the
Mott transition, are the two-body scattering
states, rather than the resonances who play a crucial role
\cite{Wergieluk:2012gd}.
Bearing in mind the technical difficulties
encountered within the present approach,
we may speculate that
one possible way to proceed while still keeping
the covariant setup, would be to put forward
the picture of complex-conjugate singularities 
in a Gribov-Zwanziger
framework, where they would be seen as elementary fields.
From a practical point of view such 
kind of modeling would use a smaller number
of fictitious states.
For example, recently it has been shown that it is possible to construct
bound states which have a Lehmann representation in the vacuum 
for a Gribov-Zwanziger model with scalar fields \cite{Capri:2012hh}.
To our best knowledge, fermionic models of such kind are
under development \cite{Dudal:2013vha}.

Alternatively, one may abandon covariant models and
use a more
physical ``gauge'', such as the Coulomb gauge, discussed {\it e.~g.} in \cite{Pak:2011wu,Watson:2012ht} for describing the 
in-medium physics
of correlations in both the hadron, and the QGP phases.
We shall come back to this question in a forthcoming investigation.

\section*{Acknowledgments}

The authors thank D. Klabu{\v c}ar for important 
contributions in the early
stages of this work and for a critical reading of the manuscript.
Discussions and comments by H.~Grigorian, T.~Hell, A.~Radzhabov,
N.~N.~Scoccola, A.~Wergieluk and D.~Zablocki are gratefully acknowledged.
S.~B. is thankful for hospitality extended to him at the JINR Dubna and at the
University of Wroc{\l}aw where much of this work was performed.
D.~H. is thankful for hospitality extended to him at the University of
Wroc{\l}aw.
S.~B. and D.~H. received support from the Ministry of Science, Education and
Sports of Croatia through the contract No. 119-0982930-1016,
D.~B. was supported in part
by the Polish National Science Centre under contract No.
DEC-2011/02/A/ST2/00306
and by the Russian Fund for Basic Research under grant number 11-02-01537-a,
while G.~C. is grateful for support by CONICET (Argentina).
This work was supported in part by CompStar, a Research Networking
Programme of the European Science Foundation and by CompStar-POL, a grant from
the Polish Ministry for Science and Higher Education supporting it.

\begin{appendix}
\section{Polarization function at finite temperature}
\label{app}
In this Appendix, the derivation of the imaginary part of the in-medium
polarization function (\ref{pol})  will be performed.
For clarity, we study the case where $\mu=0$, $\phi_3=0$,
and the mesons are at rest ${\bf q}=0$.
By analytically continuing $\omega_n\to -iz$, and using $\nu_m = -iq_0$
the integrand of the polarization function takes the following form
\be
\pi_M(z)=f^2(-z^2+\mathbf{p}^2)\frac{K_M(-z^2,\mathbf{p}^2,-q_0^2,0)}
{\mathcal{D}(-z_+^2,\mathbf{p}^2)\mathcal{D}(-z_-^2,\mathbf{p}^2)}~,
\label{int_grand}
\ee
where $z_\pm = z\pm \frac{q_0}{2}$, and where we
suppressed the $\mathbf{p}$ and $q_0$ dependence of $\pi_M$ for brevity.
Master formula for Matsubara summation is then
\be
\begin{split}
-2\pi i T\sum_{n=-\infty}^\infty
\pi_M\left(i\omega_n -\frac{i\nu_m}{2}\right) &=
\int_{i\infty}^{-i\infty} dz \pi_M(z)\\
&+\int^{i\infty+\delta}_{-i\infty +\delta}dz \pi_M(z)n(z_+)
-\int^{i\infty-\delta}_{-i\infty -\delta}dz \pi_M(z)n(-z_+)~,
\label{matsum}
\end{split}
\ee
where on the left hand side we used translational invariance, with
$n(z)=(1+e^{\beta z})^{-1}$, and $\delta>0$ infinitesimal.
It is crucial to observe that the integrals can be performed using the
information on the singularity structure of the propagator
in the whole complex plane.
Although these can be rather complicated, we shall assume that at some
not too high temperature the only singularities are simple poles at
$m_\mathrm{qp}$, see the previous discussion in the text.
Then, the only singularities of the
propagator that we need to worry about are $\pm E_\mathrm{qp}^\pm$, where
$E_\mathrm{qp} = \sqrt{v_\mathrm{qp}^2 \mathbf{p}^2+m_\mathrm{qp}^2}$,
and $E_\mathrm{qp}^a = E_\mathrm{qp}+a q_0/2$, with $a=\pm$.

Evaluating the first integral by closing the contour with a large semicircle at
$\mathrm{Re}(z)>0$ we obtain
$$\int_{i\infty}^{-i\infty} dz \pi_M(z) =
2\pi i \sum_{a=\pm}\mathrm{Res}(E_\mathrm{qp}^a)~,$$
where
\be
\mathrm{Res}(E_\mathrm{qp}^a) =
-\frac{f^2(-(E_\mathrm{qp}^a)^2+\mathbf{p}^2)}{2E_\mathrm{qp}}
\frac{K_M(-(E_\mathrm{qp}^a)^2,\mathbf{p}^2,-q_0^2,0)}
{\mathcal{D}'(-E_\mathrm{qp}^2,\mathbf{p}^2)
\mathcal{D}(-(E_\mathrm{qp} + aq_0)^2, \mathbf{p}^2)
}~.
\label{resid}
\ee
Here we denoted
\be
\mathcal{D}'(p^2) = \partial \mathcal{D}/\partial p^2~.
\label{dden}
\ee
Since the distribution function $n(z)$ has poles only on the imaginary axis,
the evaluation of the remaining integrals is performed in a similar way.
The only subtle step is acknowledging that
$n(z\pm q_0) = n(z\pm i\nu_m) = n(z)$.
For (\ref{matsum}) we obtain
\be
T\sum_{n=-\infty}^\infty \pi_M\left(i\omega_n -\frac{i\nu_m}{2}\right) =
-[1-2n(E_\mathrm{qp})]\sum_{a=\pm}\mathrm{Res}(E_\mathrm{qp}^a)~,
\label{matsum2}
\ee
where we have used that
$\mathrm{Res}(E_\mathrm{qp}^a) = -\mathrm{Res}(-E_\mathrm{qp}^a)$.

The imaginary part develops from the point where $E_\mathrm{qp}=q_0/2$
which, owing to fact that we deal with real poles,
can be handled by the $i\epsilon$ prescription.
In order to obtain (\ref{immeson}) it is sufficient to calculate
\be
\begin{split}
\mathrm{Res}(E_\mathrm{qp}^- +i\epsilon)-&
\mathrm{Res}(E_\mathrm{qp}^- -i\epsilon) =
-\frac{f^2(-(E_\mathrm{qp}^-)^2+\mathbf{p}^2)}
{2E_\mathrm{qp}}
\frac{K_M(-(E_\mathrm{qp}^-)^2,\mathbf{p}^2,-q_0^2,0)}
{\mathcal{D}'(-E_\mathrm{qp}^2,\mathbf{p}^2)}\\
&\times\left[\frac{1}
{\mathcal{D}(-(E_\mathrm{qp} - q_0+i\epsilon)^2, \mathbf{p}^2)}
-\frac{1}{\mathcal{D}(-(E_\mathrm{qp} - q_0-i\epsilon)^2, \mathbf{p}^2)}
\right]~,
\end{split}
\label{cutt}
\ee
where we have used the fact that the only discontinuities arise from the
denominator.
By expanding around $E_\mathrm{qp}=q_0/2$ ,
$$\mathcal{D}(-(E_\mathrm{qp} - q_0-i\epsilon)^2, \mathbf{p}^2)\to
2q_0 (E_\mathrm{qp}^- \mp i\epsilon)\mathcal{D}'(-q_0^2/4,\mathbf{p}^2)~,$$
and using the Plemelj formula, the following discontinuity developes
\be
\mathrm{Res}(E_\mathrm{qp}^- +i\epsilon)-
\mathrm{Res}(E_\mathrm{qp}^- -i\epsilon) =
\frac{f^2(-(E_\mathrm{qp}^-)^2+\mathbf{p}^2)}{4q_0 E_\mathrm{qp}}
\frac{K_M(-(E_\mathrm{qp}^-)^2,\mathbf{p}^2,-q_0^2,0)}
{\mathcal{D}'(-E_\mathrm{qp}^2,\mathbf{p}^2)
\mathcal{D}'(-q_0^2/4,\mathbf{p}^2)}
(-2i\pi)\delta(E_\mathrm{qp}^-)~.
\label{discc}
\ee
Plugging (\ref{discc}) into (\ref{matsum2}) and back into the original
formula (\ref{pol}) for the polarization function yields
\be
\begin{split}
\mathrm{Im}[\Pi_M(-iq_0)] = \frac{d_q}{16\pi}
&\left[1- 2n(q_0/2)\right]
\sqrt{1-\left(\frac{2m_\mathrm{qp}}{q_0}\right)^2}\\
&\times f^2\left(\frac{q_0^2}{4}-m_\mathrm{qp}^2\right)
\frac{K_M\left(0,\frac{q_0^2}{4}-m_\mathrm{qp}^2,-q_0^2,0\right)}
{\left[\mathcal{D}'
\left(-\frac{q_0^2}{4},\frac{q_0^2}{4}-m_\mathrm{qp}^2\right)\right]^2}
\theta\left(\frac{q_0}{2}-m_\mathrm{qp}\right)~.
\end{split}
\ee
Introducing the chemical potential and the Polyakov loop is now a simple
matter. By generalizing $2n(z)\to n_+^\Phi(z) + n_-^\Phi(z)$,
one arrives at (\ref{img_pol}).

\end{appendix}

\end{document}